\documentclass{rspublic}

\usepackage{verbatim}
\usepackage{graphicx}
\usepackage{amsmath,amsfonts,amssymb}
\usepackage{natbib}
\usepackage{bm}

\def\pmb#1{\setbox0=\hbox{#1}
\kern-.025em\copy0\kern-\wd0 \kern-.05em\copy0\kern-\wd0
\kern-.025em\raise.0433em\box0}
\def\bxi {{\bm \xi}}
\def\bx {{\bf x}}

\def\dint {\int\int}
\newcommand{\beq}{\begin{equation}}
\newcommand{\eeq}{\end{equation}}
\newcommand{\ba}{\begin{eqnarray}}
\newcommand{\ea}{\end{eqnarray}}

\input epsf

\usepackage[english]{babel}

\begin{document}

\title[Asymptotic theory of microstructured surfaces]{An asymptotic
  theory for  waves guided by diffraction gratings or along microstructured surfaces}
\author[T. Antonakakis, R.~V. Craster, S. Guenneau \& E.~A. Skelton]{T. Antonakakis
  $^{1,2}$, R.~V. Craster $^1$, S. Guenneau $^3$ \& E.~A. Skelton $^1$}
 \affiliation{$^1$ Department of Mathematics, Imperial College London, London SW7 2AZ, UK \\
$^2$ European Organization for Nuclear Research, CERN CH-1211, Geneva 23,
Switzerland \\
$^3$ Institut Fresnel, UMR CNRS 6133, University of Aix$-$Marseille, Marseille, France}

\label{firstpage}
\maketitle

\begin{abstract}{Plasmonics, Homogenization, Rayleigh-Bloch waves}

  An effective surface equation, that encapsulates the detail of a
  microstructure, is developed to model microstructured surfaces. The
  equations deduced accurately reproduce a key feature of surface wave
  phenomena, created by periodic geometry, that are commonly called
  Rayleigh-Bloch waves, but which also go under other names such as
  Spoof Surface Plasmon Polaritons in photonics.  Several illustrative
  examples are considered and it is shown that the theory extends to
  similar waves that propagate along gratings. Line source excitation
  is considered and an implicit long-scale wavelength is identified
  and compared to full numerical simulations. We also investigate
  non-periodic situations where a long-scale geometric variation in
  the structure is introduced and show that localised defect states
  emerge which the asymptotic theory explains.

\end{abstract}
%
\section{Introduction}
\label{sec:intro}   
It has been known for many years that surface waves, that is, waves
propagating along a surface, and exponentially decaying in amplitude
perpendicular to  the surface, are created by geometric periodic corrugations, or
perturbations, to the surface \cite[]{barlow54a,hurd54a,brekhovskikh59a}
in situations where a surface wave would otherwise not exist. Such
surface waves also exist for diffraction gratings and for trapped modes in
waveguides; these are all very similar problems
mathematically \cite[]{mciver98a,porter99a} and differ just in their setting. 
 These surface waves have been discovered in many different areas of
 wave mechanics and go under names such as edge waves \cite[]{evans93a}
 for water waves localised to periodic coastlines, spoof surface plasmon
 polaritons, SPPs,  
 \cite[]{pendry04a,fernandez11a} in modern applications of plasmonics,
 array guided surface waves \cite[]{sengupta59a} in Yagi-Uda antenna
 theory, 
 Rayleigh-Bloch surface waves \cite[]{wilcox84a,porter99a} for diffraction gratings
  amongst other areas: We will call them Rayleigh-Bloch waves as
  surface waves are typically called Rayleigh waves and Bloch waves
  arise due to periodicity. They
  can also be identified in lattice defect arrays, 
  in discrete settings \cite[]{joseph13a}, and are ubiquitous across
  wave mechanics, it is
  important to clearly delineate them from surface waves, such as Rayleigh
  waves, that are present in the absence of periodic  geometric
  features and which arise due to material mismatch or from wave mode
  coupling at the surface. 

Naturally, as these are eigenfunctions of a diffraction problem and
have implications for the uniqueness of solutions they have been the
subject of numerous existence studies
\cite[]{wilcox84a,bonnet94a,linton02a} with the conclusion that they are
a generic property of periodic surfaces and gratings that have Neumann
boundary conditions: The non-existence for Dirichlet cases is shown in \cite{wilcox84a}.

As well as being ubiquitous in wave mechanics, Rayleigh-Bloch waves are
important in applications; their dispersion characteristics can be
carefully tuned by altering only the geometry as in SPPs
\cite[]{fernandez11a}, or are important
through the coupling of incident waves into Rayleigh-Bloch waves
causing near resonant effects for finite arrays as in water waves
\cite[]{maniar97a}. These effects, and in particular the possibility to
tune or de-tune them, rely upon being able to simulate and determine
dispersion characteristics; there is advantage in being able to
represent and model them using an effective medium approach that
replaces the microstructure. 

The classical route to replace a microstructured medium with an
effective continuum representation is homogenization theory, and 
 for bulk media this is detailed in many monographs,
for instance \cite[]{sanchez80a,bakhvalov89a,bensoussan78a,panasenko05a}
and essentially relies upon the wavelength being much larger than the
microstructure which is usually assumed to be perfectly periodic:
The theory has been very versatile and has been
widely applied. 
Naturally there were extensions of this theory to
surfaces, notably by \cite{nevard97a}, again with the wavelength
limitation, unfortunately this long-wave low frequency limit is not
particularly useful at the high frequencies used in applications such
as photonics \cite[]{joannopoulos08a} and plasmonics \cite[]{maier07a,enoch12a}:
This motivated the development of high frequency homogenization (HFH) in
\cite{craster10a}. HFH breaks free of the low frequency long-wave
limitation and, for bulk media, creates effective long-scale equations
that encapsulate the microstructural behaviour, which can be upon the
same scale as the wavelength, through integrated quantities that are
no longer simple averages. The
methodology relies upon there being some basic underlying periodic
structure so that Bloch waves, and standing wave frequencies, 
encapsulate the multiple scattering between elements of the
microstructure on the short scale, and this is then modulated by a
long scale function that satisfies an anisotropic frequency dependent
partial differential equation; the technique has been successfully
applied to acoustics/ electromagnetics \cite[]{craster11a,antonakakis13a}, elastic plates  that
support bending waves \cite[]{antonakakis12a}, frames \cite[]{nolde11a}
and to discrete media \cite[]{craster10b}. The advantage of having an
effective equation for a microstructured bulk medium or surface is
that one need no longer model the detail of each individual scatterer,
as they are subsumed into a parameter on the long-scale, and attention
can then be given to the overall physics of the structure and one can
identify, or design for, novel physics. 

The HFH theory of \cite{craster10a} is not alone: There is
considerable interest in creating effective continuum models of
microstructured media, in various related fields, that break free from
the conventional low frequency homogenisation limitations. This desire
has created a suite of extended homogenization theories originating in
applied analysis, for periodic media, called Bloch homogenisation
\cite[]{conca95a,allaire05a,birman06a,hoefer11a}. There is also a
flourishing literature on developing homogenized elastic media, with
frequency dependent effective parameters, also based upon periodic
media as in \cite{nematnasser11a}. Those approaches notwithstanding, our aim
here is to extend the HFH theory to microstructured surfaces and
obtain frequency dependent effective surface conditions that capture
the main features of the surface waves that exist.

Our aim herein is to generate a surface HFH theory for structured surfaces
in the context of perfect surfaces, importantly one can  modify the
theory, as done for bulk waves in \cite{craster11a,antonakakis12a,makwana13a}, to pull out
defect states associated with non-periodic variation. It is also
important to note that the HFH theory has a deep connection with the
high frequency long wavelength near cutoff theory of waveguides
\cite[]{craster13a} and the defect states are related to localisation by deformed
waveguides \cite[]{gridin04a,kaplunov05a,gridin05a}. We also
naturally extend the HFH theory to diffraction gratings. 
In section
\ref{sec:general} the theory is created culminating in the
effective equation that encapsulates the surface
behaviour. Illustrative examples, in section \ref{sec:examples}, then
show the efficacy of the methodology versus the dispersion relations
found numerically. An interesting practical situation is where some
geometric variation occurs, then one expects the possibility of
trapped modes along the structure occuring at a set of discrete frequencies,
and we consider a comb-like structure where the teeth have varying length
in section \ref{sec:defectState}; the asymptotic theory is compared to
full numerical simulations. Finally, concluding comments and remarks
are drawn together in section \ref{sec:conclude}. 

\begin{figure}
  \begin{center}
    \includegraphics[scale=0.9]{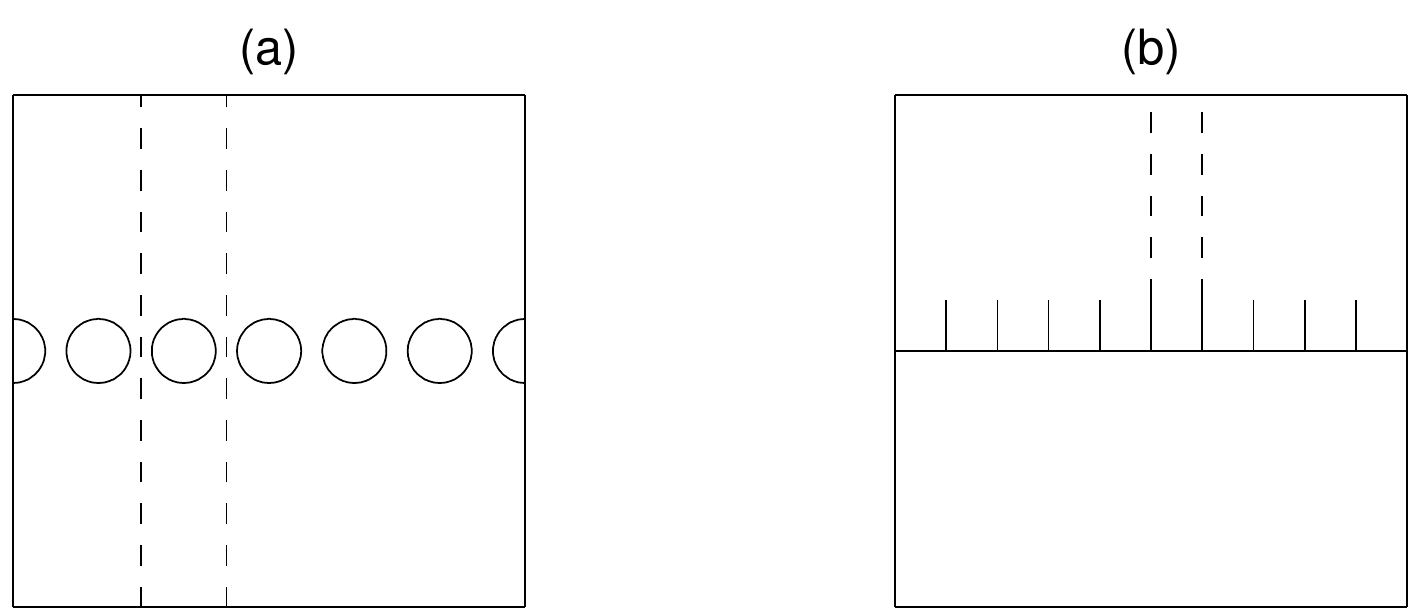}
   \end{center}
\caption{A diffraction grating of cylinders shown in (a), and (b)
  shows a periodic ``comb'' surface that supports spoof surface
  plasmons. Both panels have the 
  elementary strip shown as the dashed lines.}
\label{fig:schematic}
\end{figure}

\section{General theory}
\label{sec:general}
For perfect infinite linear arrays, diffraction gratings or surface
structures arranged
periodically, one focuses attention on a single elementary strip of
material that then repeats (see Fig. \ref{fig:schematic} for
illustrative cases);
quasi-periodic Floquet-Bloch boundary conditions describe the
phase-shift across the strip as a wave moves from strip to strip through the material. Rayleigh-Bloch
waves are special as they consist of waves that also decay
exponentially in the perpendicular direction away from the array. 
 Dispersion
relations  then relate the Floquet-Bloch wavenumber, the
phase-shift, to frequency. Although the problem is truly
two-dimensional, the assumption of exponential decay in the 
perpendicular renders it quasi-one dimensional with the wavenumber
remaining scalar; this contrasts with the theory of Bloch waves in
photonic crystals \cite[]{joannopoulos08a} where a vector wavenumber and
the Brillouin zone are more natural descriptions.

We shall approach the problem tangentially and generate an asymptotic
theory, importantly we take Neumann boundary conditions on the lattice or
surface; physically,  this can be considered as TE (transverse electric) polarization for a perfectly
conducting surface which is a good model for microwaves \cite[]{petit80a}. 

A time harmonic dependence of propagation $\exp(-i\omega t)$, with
 frequency $\omega$, is
assumed throughout, and henceforth suppressed, 
 and after non-dimensionalisation one arrives at 
\beq
\nabla^2_{\bx}u(\bx)+\Omega^2 u(\bx)=0,
\qquad
\text{with} \quad \Omega = \frac{\omega l}{c},
\label{eq:2dnormal}
\eeq
 where $l$ is the lengthscale of the micro-scale and $c$ is the
 wavespeed, as the governing equation of interest. We consider the half-space 
 $-\infty<x_1<\infty$, $0<x_2<\infty$, and for the grating extend to
 $-\infty$ in $x_2$. In (\ref{eq:2dnormal}), $\Omega$ is the non-dimensional frequency
and $u$ is the out-of-plane displacement in elasticity or the $H_3$
component of the magnetic field in TE polarisation. 

The two-scale nature of the problem is incorporated using small
and large length scales to define two new independent coordinates
namely $ X=x_1/L$, and $(\xi_1,\xi_2)=(x_1,x_2)/l$. The implicit
assumption is that there is a small scale, characterized by $l$, and a
long scale characterized by $L$ where $\epsilon=l/L\ll 1$. As the
structure is quasi-one dimensional, with the mismatch in the scales
being just along the structure, we only introduce a single long-scaled
variable in $X$; we do not introduce a long-scale $Y$ in the $x_2$
direction as it is redundant. 

Under this rescaling, equation
(\ref{eq:2dnormal}) then becomes,
\beq
\left[
\frac{\partial^2}{\partial
  \xi_1^2}+2\epsilon\frac{\partial^2}{\partial \xi_1\partial 
  X}+\epsilon^2\frac{\partial^2}{\partial
  X^2}+\frac{\partial^2}{\partial \xi_2^2}+\Omega^2
\right]
 u(X,\xi_1,\xi_2)=0.
\label{eq:two-scales2D}
\eeq
Standing waves, that exponentially decay perpendicular to the surface/
grating, can occur when there are periodic (or anti-periodic)
boundary conditions across the elementary strip (in the $\bxi$
coordinates) and these standing waves encode the local information
about the multiple scattering that occurs by the neighbouring
 strips. The asymptotic technique we create is a perturbation about these
standing wave solutions, as these are associated with 
 periodic and anti-periodic boundary conditions, which are
 respectively in-phase and out-of-phase waves across the strip, the
 conditions on the short-scale $\bxi$ on the edges of the strip, $\partial S_1$, are known: 
\beq
u|_{\xi_{1}=1}=\pm u|_{\xi_{1}=-1} \quad \text{and} \quad u_{,\xi_1}|_{\xi_{1}=1}=\pm u_{,\xi_1}|_{\xi_{1}=-1},
\label{eq:periodicBC}
\eeq
 where $u_{,\xi_i}$ denotes differentiation of $u$ with respect to variable $\xi_i$ and 
 with the $+,-$ for periodic or anti-periodic cases
 respectively. There is therefore a local solution on the small scale
 that incorporates the multiple scattering of a periodic medium and
 that will then be modulated by a long-scale function that satisfies a
 differential equation. Typically, the periodic case corresponds to long-waves relative to
 the structure - this case is not particularly interesting and is
 captured by conventional low-frequency homogenisation. We therefore
 concentrate upon the anti-periodic case.

  We pose an ansatz for the field and the frequency, 
\ba
u( X,\bxi)=u_0(X,\bxi)+\epsilon u_1(X,\bxi)+\epsilon^2 u_2(X,\bxi)+\ldots , 
\nonumber\\ 
\Omega^2=\Omega_0^2+\epsilon \Omega_1^2+\epsilon^2 \Omega_2^2+\ldots
\label{eq:expansion2D}
\ea
The $u_i(X,\bxi)$'s adopt the boundary conditions
(\ref{eq:periodicBC}) on the short-scale, with the minus sign for anti-periodicity, on
the edge of the strip.  An ordered hierarchy of 
equations emerge in powers
of $\epsilon$, and are treated in turn
\beq
u_{0,\xi_i\xi_i} + \Omega_0^2 u_0=0,
\label{eq:leadingOrder}
\eeq
\beq
u_{1,\xi_i\xi_i} + \Omega_0^2 u_1=
-2 u_{0,\xi_1 X}
-\Omega_1^2 u_0,
\label{eq:firstOrder}
\eeq
\beq
  u_{2,\xi_i\xi_i} + \Omega_0^2 u_2 =-u_{0,XX}
   -2u_{1,\xi_1 X} 
-\Omega_1^2 u_1 -\Omega_2^2 u_0.
  \label{eq:secondOrder}
\eeq
The leading order equation (\ref{eq:leadingOrder}) is independent of
the long-scale $X$ and is a standing wave on the elementary strip
existing at a specific
eigenfrequency $\Omega_0$ and has associated eigenmode
$U_0(\bxi;\Omega_0)$, modulated by a long-scale function $f_0(  X)$
and so we expect to get an ordinary differential equation (ODE) for $f_0$ as
an effective boundary, or interface, condition characterising
the grating when viewed from afar: To leading order 
\beq
 u_0(X,\bxi)=f_0(X) U_0(\bxi;\Omega_0).  
\label{eq:LeadingSolution}
\eeq 
The entire aim is to arrive
at an ODE for $f_0$ posed entirely upon the long-scale, but with the
microscale incorporated through coefficients that are integrated, not
necessarily averaged, quantities. 

Before we continue to next order, equation (\ref{eq:firstOrder}), we
define the Neumann boundary conditions on the inclusions, or the
micro-structured surface, $\partial S_2$, as
\beq
\frac {\partial u} {\partial {\bf n}} =u_{,x_i} n_i|_{\partial S_2}=0,
\label{eq:neumann}
\eeq
 using Einstein's notation for summation over repeated indices, and 
 where ${\bf n}$ is the outward pointing normal, 
 which in terms of the two-scales and $u_i( X, \bxi)$ become
\beq
U_{0,\xi_i}n_i = 0,\quad
\label{eq:1rstNeumann}
U_0f_{0,X}n_1 + u_{1,\xi_i}n_i=0,\quad
u_{1,X}n_1 + u_{2,\xi_i} n_i=0.
\eeq
The leading order eigenfunction $U_0(\bxi;\Omega_0)$ must satisfy the
first of these conditions and it is relatively straightforward to
extract this either numerically, as we do later, or using
semi-analytic methods such as the  residue calculus technique \cite[]{hurd54a}. 

Moving to the first order equation
(\ref{eq:firstOrder}) we invoke a solvability condition by integrating
over the elementary strip, which is on the short-scale ${\bm \xi}$, the product of equation (\ref{eq:firstOrder}) and $U_0$
minus the product of equation (\ref{eq:leadingOrder}) and
$u_1/f_0({X})$: The result is that the 
eigenvalue $\Omega _1$ is identically zero. 

We then solve for
$u_1=f_{0,X}U_{1}(\bxi)$, so $ U_1$ satisfies
\beq
    \nabla_{\bxi}^2 U_1+ \Omega_0^2 U_1=-2 U_{0,\xi_1}
\eeq
 subject to the boundary condition
\beq
 {\bf n}\cdot \nabla_{\bxi} U_1=-U_0 n_1,
\eeq
 on $\partial S_2$. Again solutions can be found numerically or using semi-analytic methods.

Going to second-order, 
 a similar solvability condition to that used at first order is
 applied using equation
(\ref{eq:secondOrder}); after some algebra we obtain the desired ordinary differential equation for $f_0$ 
\ba
T f_{0,XX}+\Omega_2^2f_0=0
\label{eq:f_0}
\ea
posed entirely on the long-scale $X$. The coefficient $T$ is
constructed from integrals over the elementary strip in ${\bxi}$ and is ultimately
independent of $\bxi$.  The formula for $T$ is 
\beq
T {\dint_S U_0^2dS}=\dint_S (U_0^2 +2U_{1,\xi_1}U_0) dS-\int_{\partial S_2}U_1U_0n_1ds, 
\label{eq:t11}
\eeq
 which using Green's Theorem, with vector field ${\bf F}=(U_1U_0,0)$, 
  simplifies to,
\beq
T {\dint_S U_0^2dS}=\dint_S (U_0^2 +U_{1,\xi_1}U_0-U_{0,\xi_1}U_1) dS.
\label{eq:t11simple}
\eeq
For an infinite perfect grating, the Bloch grating, one sets
$f_0(X)=\exp(i[\pi/2-\kappa] X/\epsilon)$ and equation (\ref{eq:f_0})
simplifies to $\Omega_2^2=[\pi/2-\kappa]^2T/\epsilon^2$ and from
(\ref{eq:expansion2D}) the asymptotic dispersion relation relating
frequency, $\Omega$, to Bloch wavenumber, $\kappa$, is
\beq
\Omega\sim\Omega_0+\frac{T}{2\Omega_0}[\pi/2-\kappa]^2.
\label{eq:asyEqu}
\eeq 
 In (\ref{eq:asyEqu}) $T$ is invariably negative, c.f Table
 \ref{tab:first_four} for some illustrative values, as is $\Omega_2^2$, the latter should
 not be confused with having negative frequencies as it is just a
  frequency perturbation.
Therefore, if the surface or grating supports Rayleigh-Bloch waves
then they are represented as an effective string or membrane
equation (\ref{eq:f_0}) where the effective stiffness (or effective
inverse of permittivity in the context of photonics) of the string is
$T$; all the microstructural and geometrical information is contained 
in this asymptotic result and one can then extend it to be used for finite
arrays or for slightly non-periodic arrays, or forced problems etc,
but our aim here is to now demonstrate that this theory is well-founded.

\begin{figure}
\centering
\includegraphics[scale=0.7]{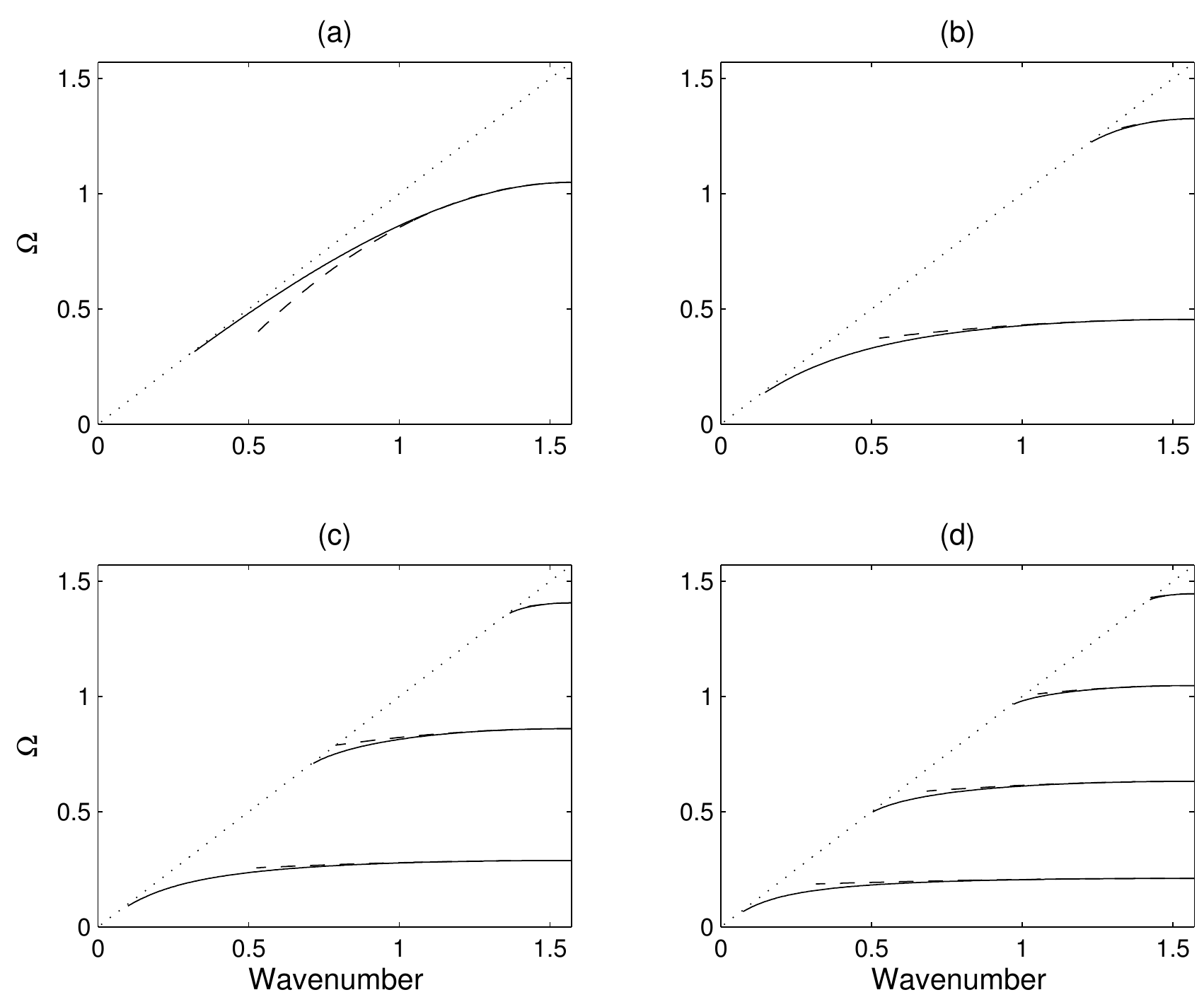}
 \caption{The dispersion branches for the comb-like structure.  
   Solid-lines are from (\ref{eq:hurd}), and the light-line
   $\Omega=\kappa$ is 
   the dotted line. Panels (a), (b), (c) and (d) are 
    for $a=1$,
   $3$, $5$ and $7$ respectively. Asymptotics from HFH
  (\ref{eq:asyEqu}), with $T$ given in Table
   \ref{tab:first_four} for $a=7$ (d), are shown as dashed curves. 
}
\label{fig:comb}
\end{figure}

\begin{figure}
\centering
\includegraphics[scale=0.7]{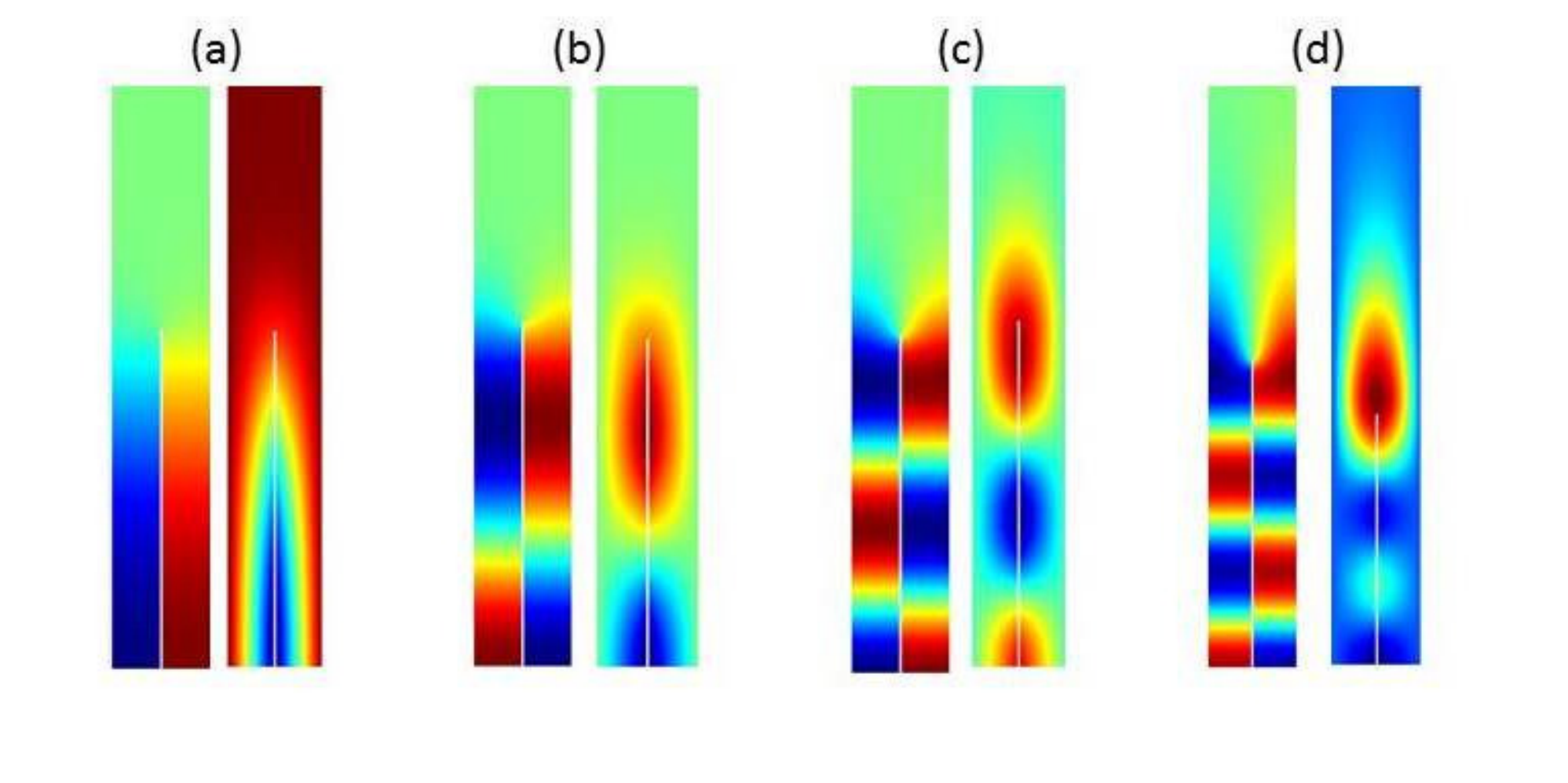}
 \caption{The eigenfunctions $U_0$, and $U_1$ shown for the
   comb-like structure with $a=7$ c.f. Fig. \ref{fig:comb}(d). These are for the standing wave
   frequencies $\Omega_0$ in Table \ref{tab:first_four} with (a)-(d)
   for ascending $\Omega_0$. In each panel $U_0$ is shown on the left
   and $U_1$ on the right.}
\label{fig:EigenComb}
\end{figure}

\begin{table}
\centering
\begin{tabular}{cc}\hline
$T$ & $ \Omega_0$\\
$ -0.006485497624108$ & $0.210161050669707 $ \\
$ -0.067470169867289$ & $0.629209426388598 $ \\
$ -0.280897588912595$ & $1.043323585456635 $ \\
$ -2.350025233704123$ & $1.440535862845912 $ \\
\hline
\end{tabular}

\caption{The four standing wave frequencies for the comb-like structure with $a=7$,
  c.f. Fig. \ref{fig:comb}(d), together with associated values for
  $T$.}
\label{tab:first_four}
\end{table}

\begin{figure}
\centering
\includegraphics[scale=0.2]{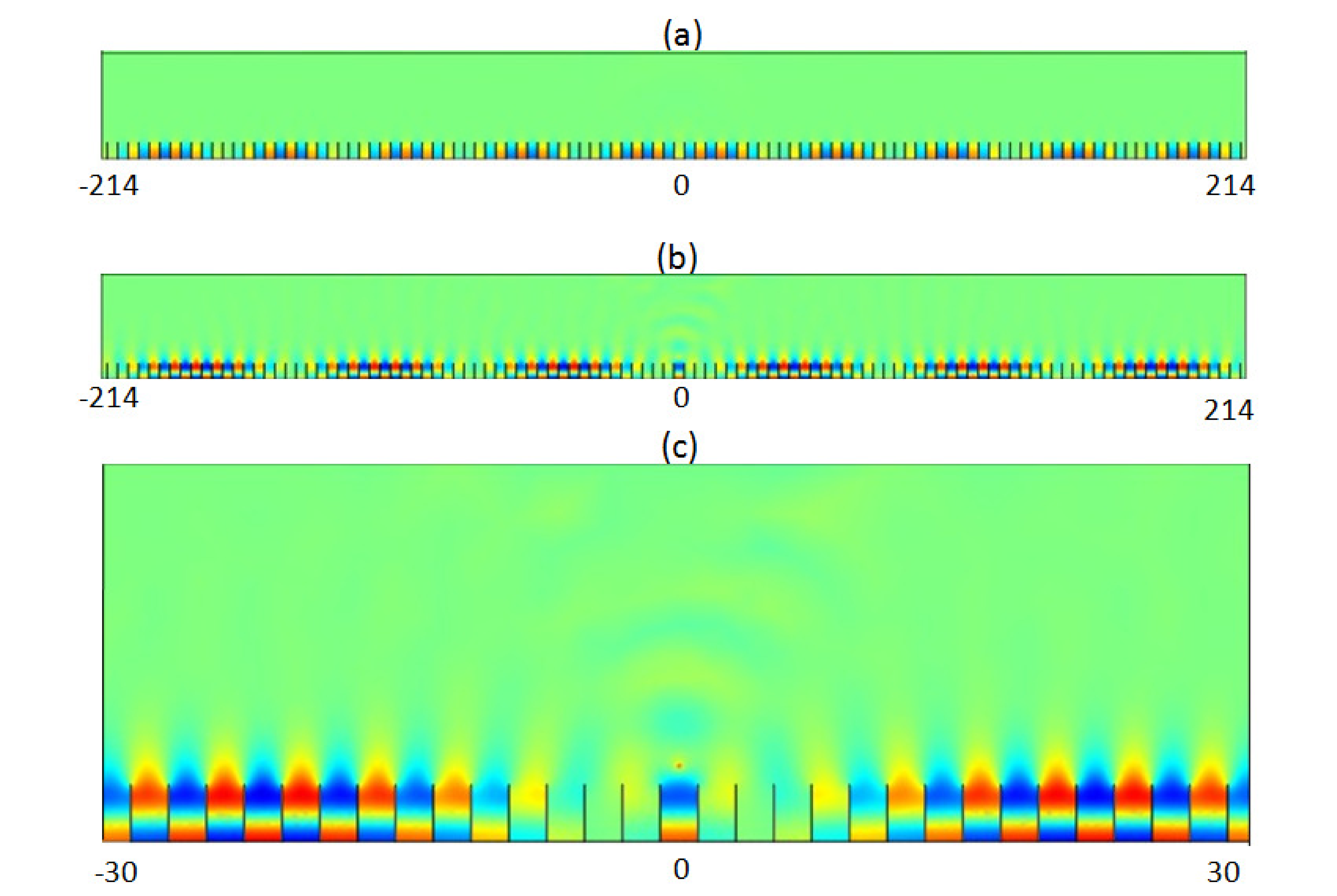}
 \caption{Plots of real$(u)$ from finite element simulations for a comb grating with $a=3$ c.f. Fig. \ref{fig:comb}(b):
(a) Fields generated by a line source with
$\Omega=0.4509$ ($\Omega_0=0.45127$); (b) Fields generated by a line source with
 $\Omega=1.3138$ ($\Omega_0=1.31510$);(c) Detail of real $(u)$ at $\Omega=1.3138$.} 
\label{fig:seb2}
\end{figure}

\begin{figure}
\centering
\includegraphics[scale=0.85]{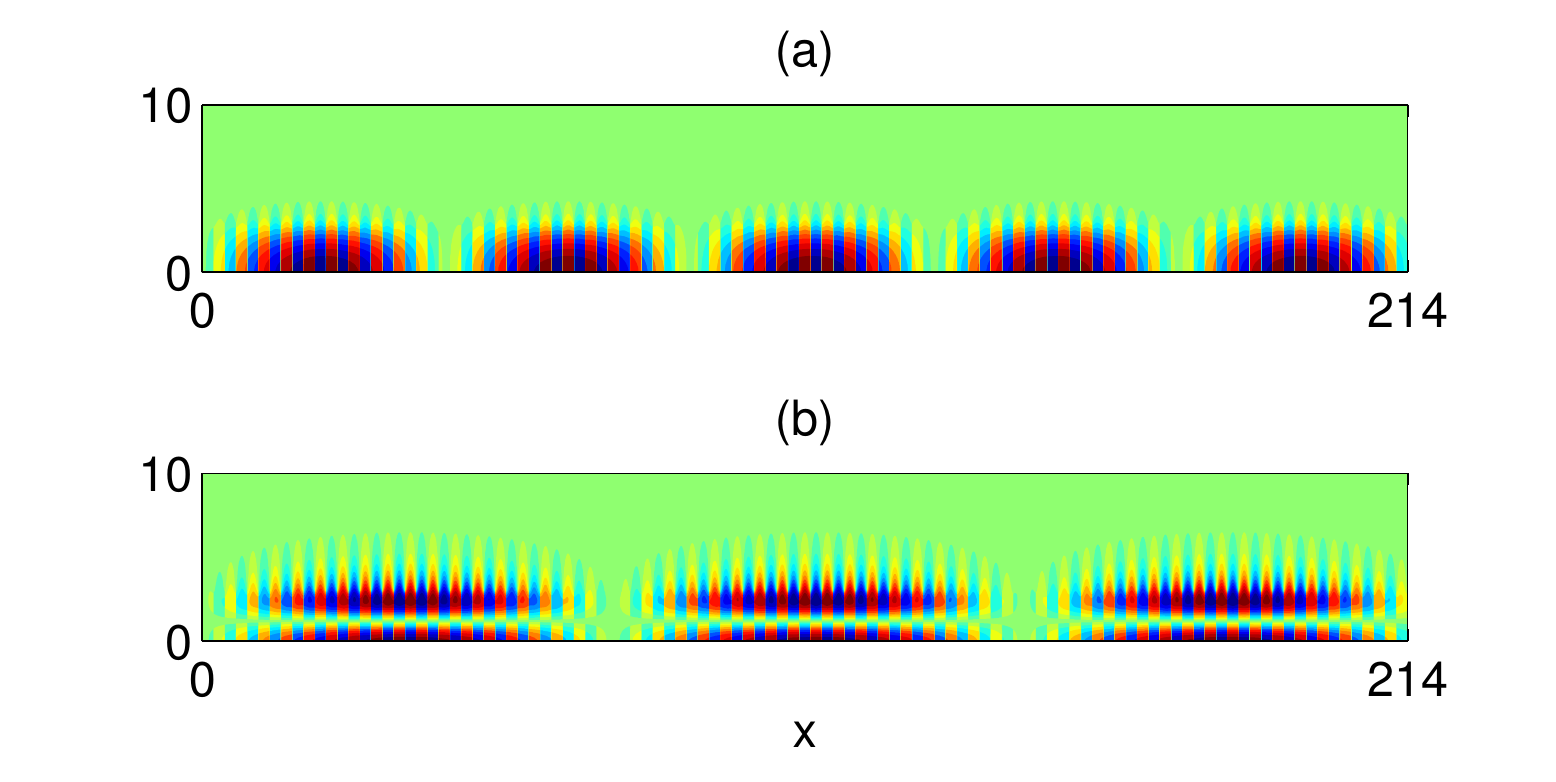}
 \caption{Plots of real$(u)$ from HFH for a comb with $a=3$, 
   c.f. Fig. \ref{fig:seb2},
 generated by a line source at (a) $\Omega=0.4509$ and  (b)
 $\Omega=1.3138$ respectively.}
\label{fig:effComb}
\end{figure}

\subsection{The classical long wave zero frequency limit}
\label{sec:classical}
The current theory simplifies dramatically in the classical long wave,
low frequency, limit where $\Omega^2\sim O(\epsilon^2)$, this
is a periodic case on the short-scale:
  $U_0$ becomes uniform, and without loss of generality, is set to be
 unity over the elementary strip. The final equation is again
 (\ref{eq:f_0}) where  $T$ simplifies to 
\beq
T\dint_S dS=\dint_S dS+\dint_S U_{1,\xi_1}dS.\quad 
\label{eq:tij0}
\eeq
 $U_{1_i}$ satisfies the Laplacian \(
U_{1,\xi_i\xi_i} =0
\)
 and  $U_{1}$ has boundary conditions 
\(
U_{1,\xi_i}n_i=-n_1
\label{eq:NeumannU1}
\)
on $\partial S_2$.
Rearranging equation (\ref{eq:tij0}) yields, using a rectangular
strip of height $y^*$ for $S$, 
\beq
T=1+\frac{\dint_S
  U_{1,\xi_i}dS}{\lim\limits_{y^*\to+\infty}\int_0^{y^*}2
  dy}\rightarrow 1,
\label{eq:Tij}
\eeq
 from which $\Omega=\kappa$ and thus the light-line of unit slope
 emerging from the origin arises asymptotically.

\begin{figure}
\centering
\includegraphics[scale=0.7]{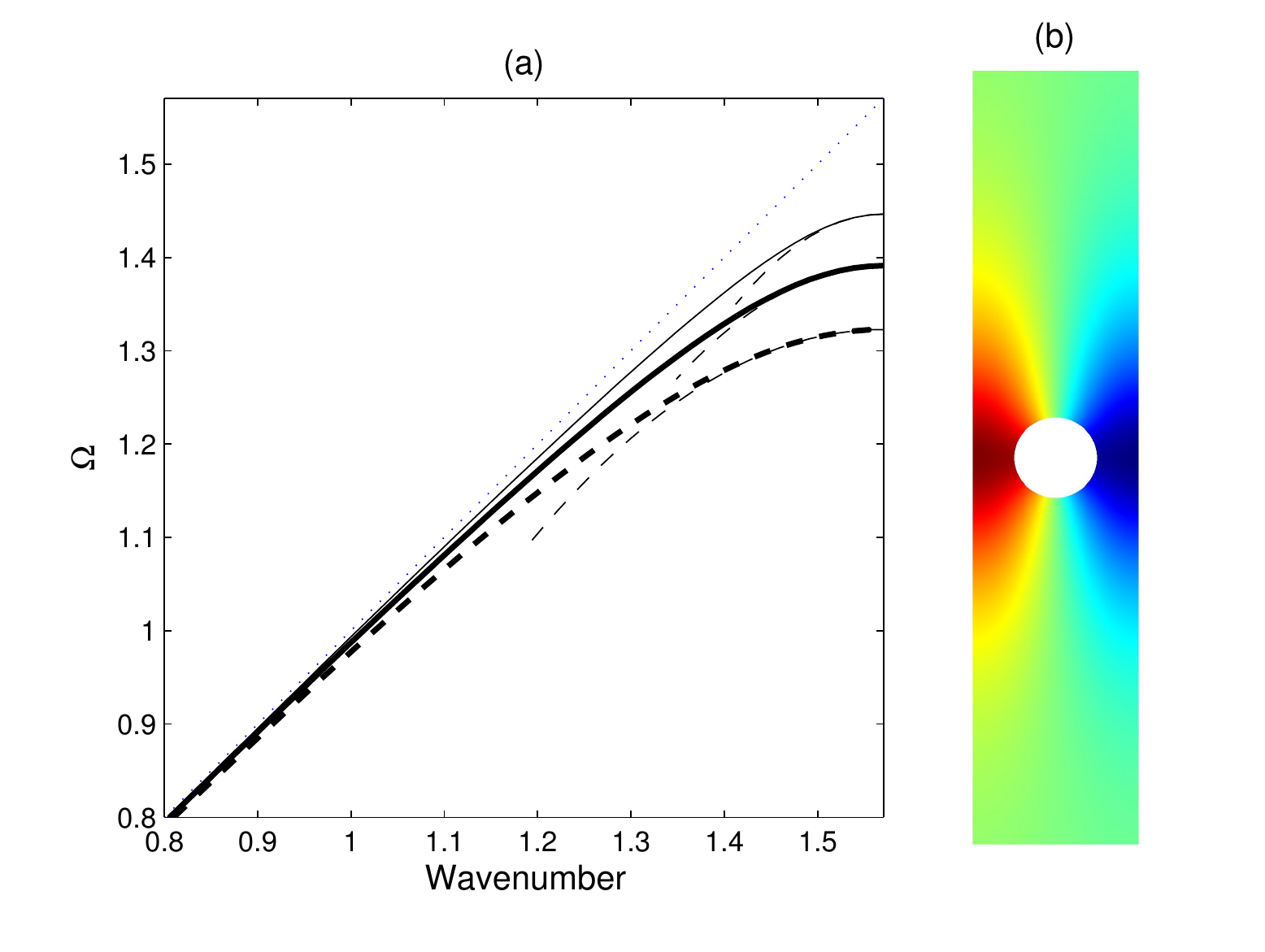}
 \caption{The dispersion branches, for the symmetric mode, 
   ($\Omega=\kappa$ as dotted line) are shown in panel (a)
   for cylinders with radii of
   $0.7$ (dashed bold), $0.5$ (bold) and $0.4$ (solid). The corresponding
   asymptotic curves, from  (\ref{eq:asyEqu}),  are shown as dashed lines. Panel (b) shows the standing wave eigensolution $U_0$ for $r=0.5$.
}
\label{fig:circleSymmetric}
\end{figure}

\begin{figure}
\centering
\includegraphics[scale=0.7]{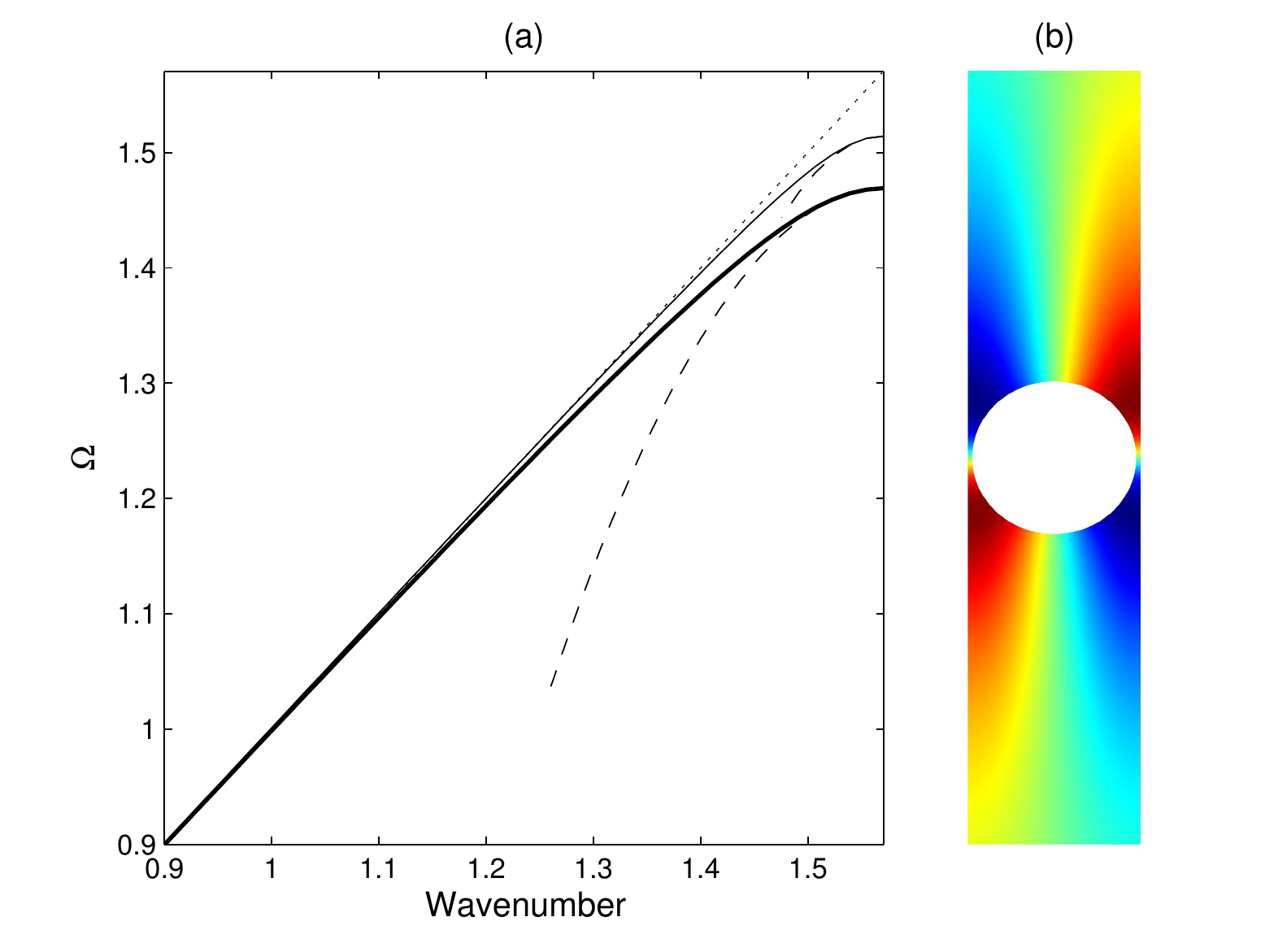}
 \caption{The dispersion branches, for  the antisymmetric mode, 
   ($\Omega=\kappa$ as dotted line)
  are shown in panel (a)  for cylinders with radii of
  $0.95$ (solid line) and $0.99$ (bold solid line).
The asymptotics are shown as dashed lines. In panel (b) the antisymmetric eigensolution
  $U_0$ is shown for $r=0.95$.
}
\label{fig:circleAsymmetric}
\end{figure}

\begin{figure}
\centering
\includegraphics[scale=0.5]{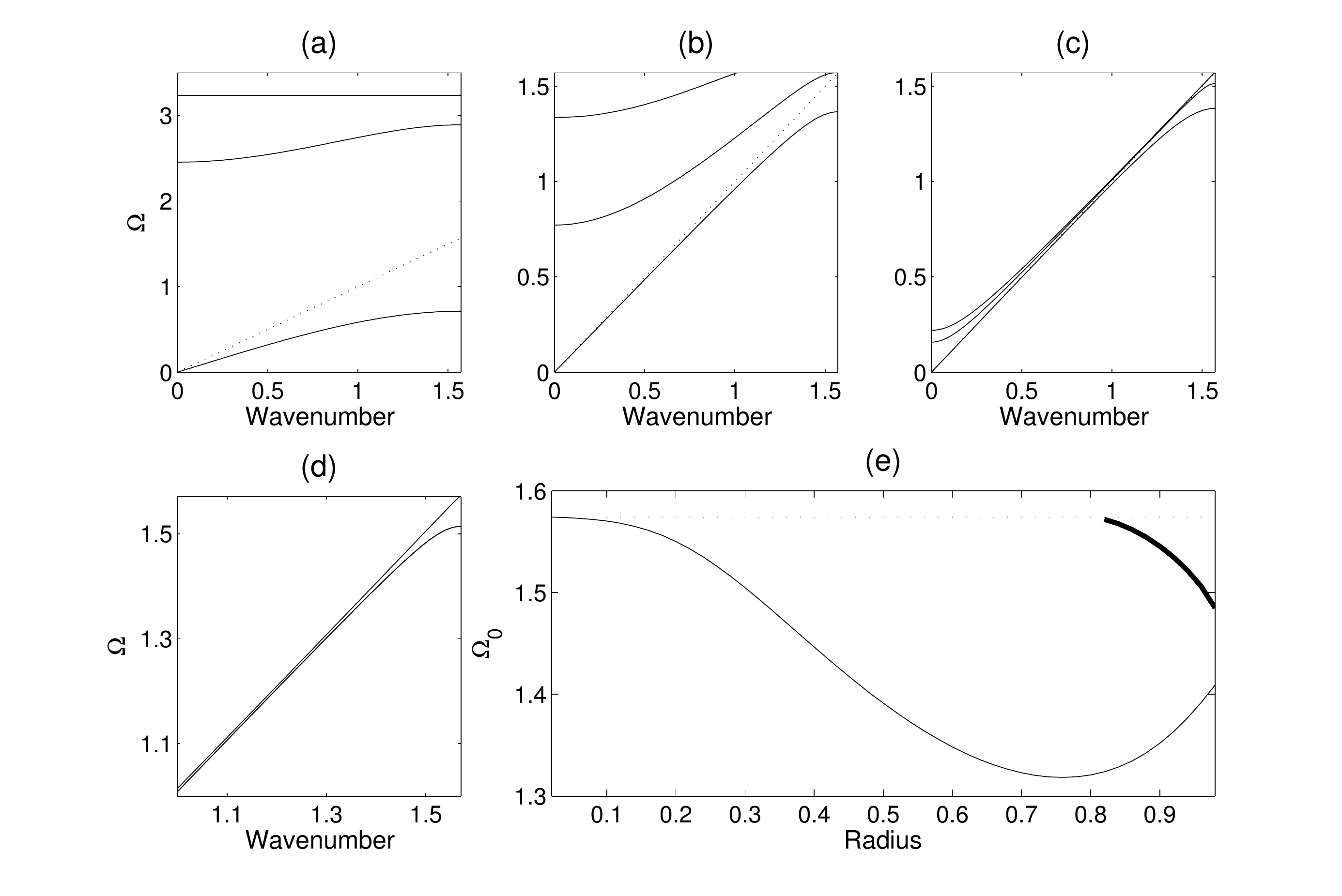}
 \caption{The dispersion curves for cylinders placed in a rectangular
   array are shown for Bloch waves in the $x_1$ direction. 
The lowest three dispersion branches, as solid lines, and
$\Omega=\kappa$ as dotted, are shown in (a), (b) and (c) for rectangle
heights of $2$, $6$ and $30$ respectively where the cylinder radius is
$0.95$. Panel
(d) shows the dispersion curves for a cell of height $30$, in solid
the second modes for the respective radii of $0.4$ and $0.95$
and the lightline in dashed. 
Panel (e) shows the variation of standing wave frequencies for the
symmetric (solid) and asymmetric (bold) modes versus cylinder 
 radius for the infinite strip.}
\label{fig:Parametriccircle}
\end{figure}

\begin{figure}
\centering
\includegraphics[scale=0.175]{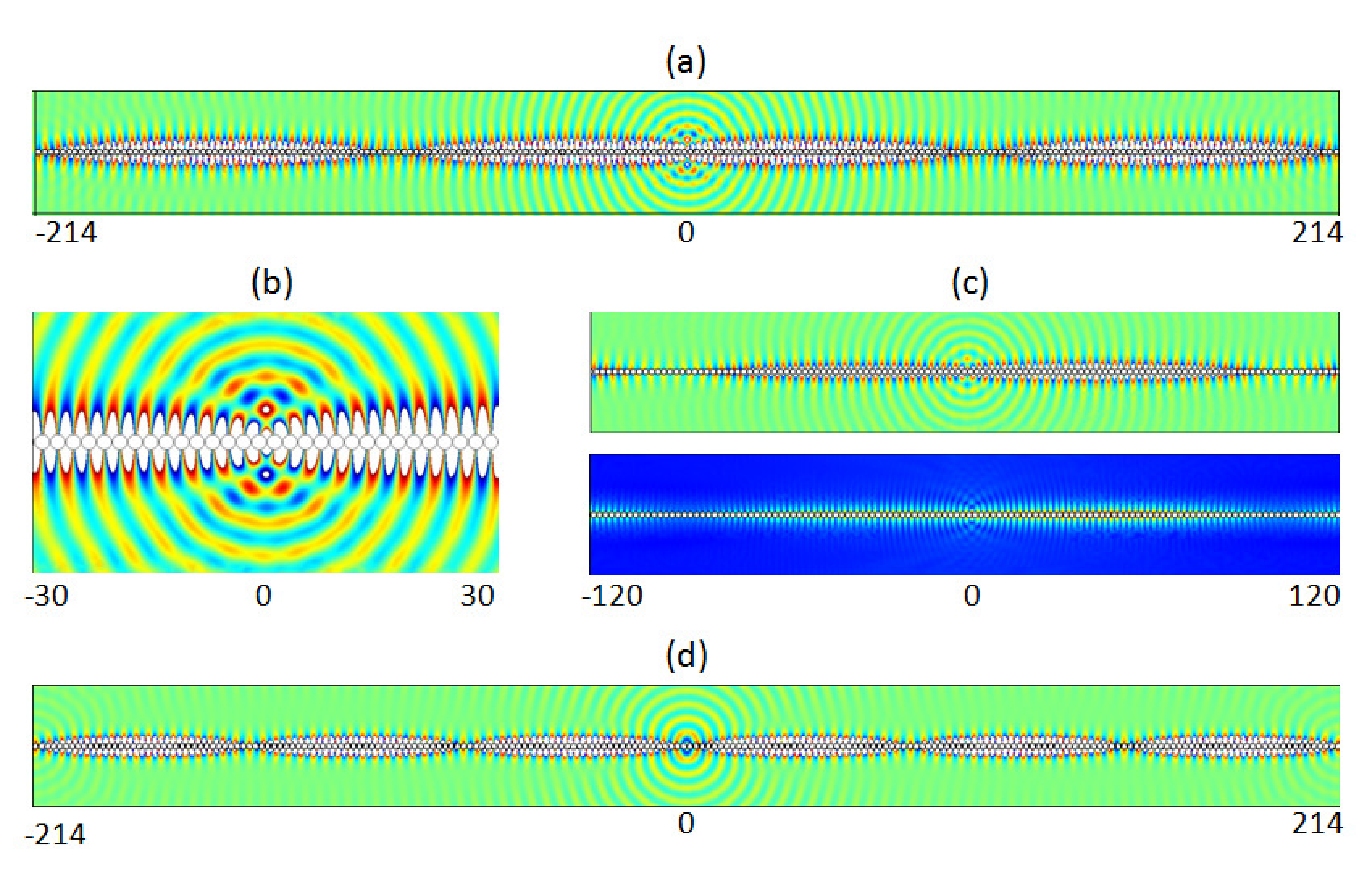}
 \caption{Plots of real $(u)$ for a diffraction grating consisting of
   cylinders of radius $r=0.95$: (a) Asymmetric fields generated by a
   dipole source for $\Omega=1.508$ ($\Omega_0=1.51445$); (b) Detail
   close to the dipole source showing the micro-scale asymmetry; (c)
   Detail of the real part (upper panel) and absolute value (lower
   panel) of the asymmetric $u$ at $\Omega=1.508$; (d) Symmetric
   fields generated by a line source at frequency $\Omega=1.38$
   ($\Omega_0=1.38407$).}
\label{fig:seb1}
\end{figure}

\begin{figure}
\centering
\includegraphics[scale=0.85]{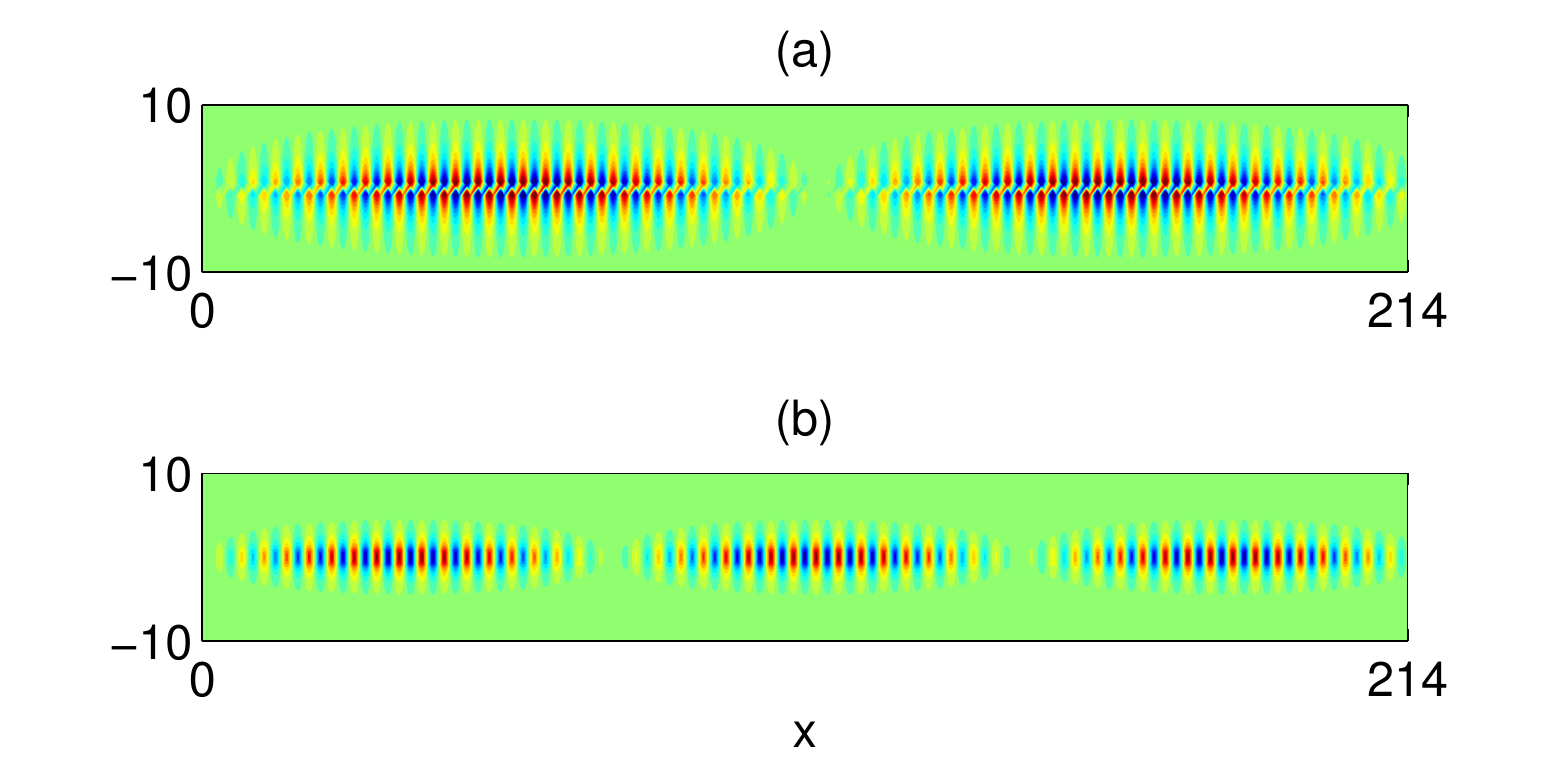}
 \caption{Plots of real $(u)$ from HFH for a cylinder of radius $r=0.95$, 
   c.f. Fig. \ref{fig:seb1},
 generated by a line source at (a) $\Omega=1.508$ and (b) $\Omega=1.38$ respectively}
\label{fig:effCircle}
\end{figure}

\subsection{A dynamic characteristic length scale}
\label{sec:effMedia}
As we will see later, in section \ref{sec:examples}, equation
(\ref{eq:asyEqu}) is an excellent asymptotic approximation for the
dispersion diagrams of such gratings which verifies the validity of
HFH. Ultimately one wishes to homogenize a periodic, or nearly
periodic, surface and this is achieved with equation (\ref{eq:f_0}) transformed back in the original coordinates together with the replacement of $\Omega_2$ using the asymptotic expansion in equation (\ref{eq:expansion2D}). The effective medium equation resulting from such operations is,
\beq
Tf_{0,xx}+(\Omega^2-\Omega_0^2)f_0=0.
\label{eq:effEqu}
\eeq
 The solutions of equation (\ref{eq:effEqu}) are harmonic with argument
 $\sqrt{(\Omega^2-\Omega_0^2)/{T}}$ provided $\Omega<\Omega_0$ and
 $T<0$. It is now clear that if the excitation frequency is slightly
 away from the standing wave frequency there will be an oscillation
 with wavelength $\lambda=2\pi\sqrt{T/(\Omega^2-\Omega_0^2)}$ which
 will represent a characteristic length scale for such an infinite
 periodic medium. That length scale not only depends on the excitation
 and standing wave frequencies but also on the homogenized parameter 
 $T$ that represents dynamically averaged material
 parameters. Therefore one observes highly oscillatory behaviour with
 each neighbouring strip out-of-phase but modulated by a long-scale
 oscillation of wavelength $\lambda/2$ reminiscent of a beat frequency
 but induced by the microstructure; the $\lambda/2$ arises as one
 observes both $f_0$ and $-f_0$.  

\section{Illustrative examples}
\label{sec:examples}
We now illustrate the theory using linear arrays of cylinders, split ring resonators (SRR) and a
comb-like surface structure as these are exemplars of the situations
seen in practice. 

\subsection{The classical comb}
\label{sec:comb}
An early example for which Rayleigh-Bloch waves were found explicitly
is that of a Neumann comb-like surface consisting of periodic thin
plates of finite length, $a$, perpendicular to a flat wall and distant by $2l$ from each other. This was
initially studied by \cite{hurd54a} with later modifications by
\cite{desanto72a,evans93a,evans02a}: 
 It is a canonical example and can be
considered as a diffraction grating if extended to the negative
half-plane by reflexion symmetry. 

We will concentrate upon non-embedded Rayleigh-Bloch waves in
$\Omega<\kappa$ and Hurd's dispersion relation 
\beq
 \Omega a/l=(n+1/2)\pi+2\Omega/\pi\ln 2 +\chi(\kappa,2\Omega)
\eeq
 where
\beq
 \chi(\kappa,\Omega)=-\sin^{-1}(\Omega/\kappa)+\sum_{n=1}^\infty \left(
\sin^{-1}(\Omega/n\pi)- \sin^{-1}(\Omega/(\kappa+2n\pi)) 
-\sin^{-1}(\Omega/\vert\kappa-2n\pi\vert)\right)
\label{eq:hurd}
\eeq
 provides a highly accurate approximation; dispersion branches are shown in
Fig. \ref{fig:comb} using Hurd's formulae. There exists an even more accurate
result from \cite{evans93a} which is virtually indistinguishable from
that of Hurd, and it is possible, as we also do here, to use finite
elements to model the comb numerically, the only detail of note is
that the comb teeth have finite width of $0.05$ in the finite element
simulations to avoid any numerical issues at the tip of the teeth, and all these methods give
coincident results. The dispersion equations (\ref{eq:hurd}) come from
a Fourier series approach and we also investigate this approach
numerically and provide results in Tables
\ref{tab:ES1},\ref{tab:ES2}. 
The parameter $a$ is the length of the
tooth, and the curves are locally quadratic near $\pi$ as we
expect from (\ref{eq:asyEqu}); clearly the HFH asymptotics provide an
excellent representation of the dispersion curves close to the
standing wave frequency as illustrated by the dashed curves in
Fig. \ref{fig:comb}. The standing wave frequencies $\Omega_0$ and the
effective parameter $T$ are given in Table \ref{tab:first_four} for
the case $a=7$. 
Increasing $a$ corresponds to more dispersion curves appearing and the
eigensolutions for $a=7$ are shown in 
Fig. \ref{fig:EigenComb} together with their $U_1$ counterparts, and the reason for the
increasing number of surface modes is immediately apparent being
intimately connected with the number of modes the open waveguide supports. The $U_0$ modes decay rapidly as
they exit the open waveguide particularly for the lowest standing wave
frequencies. 

\begin{figure}
\hspace{3cm}(a)\hspace{6.5cm}(b)
\includegraphics[scale=0.45]{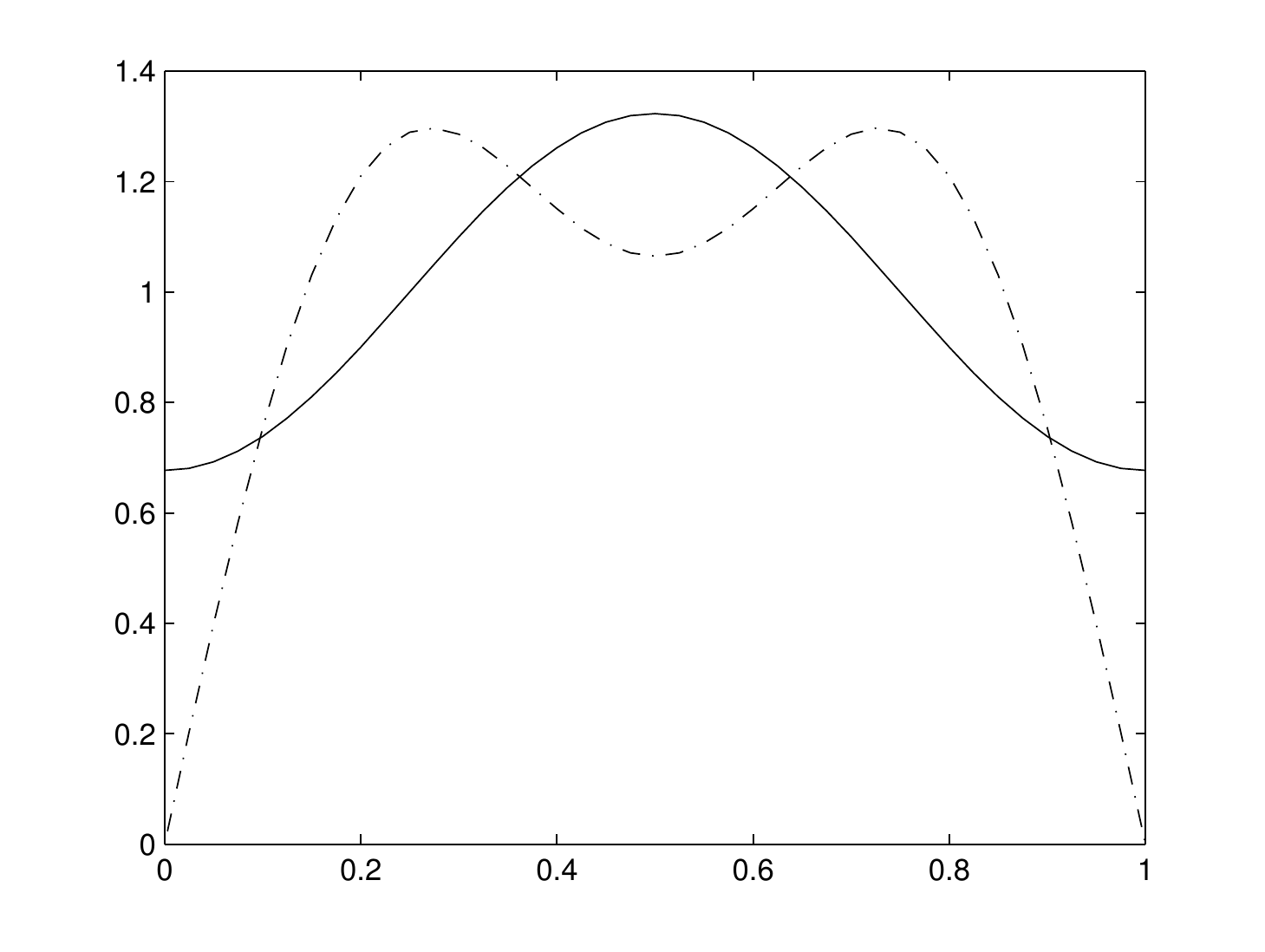} 
\includegraphics[scale=0.45]{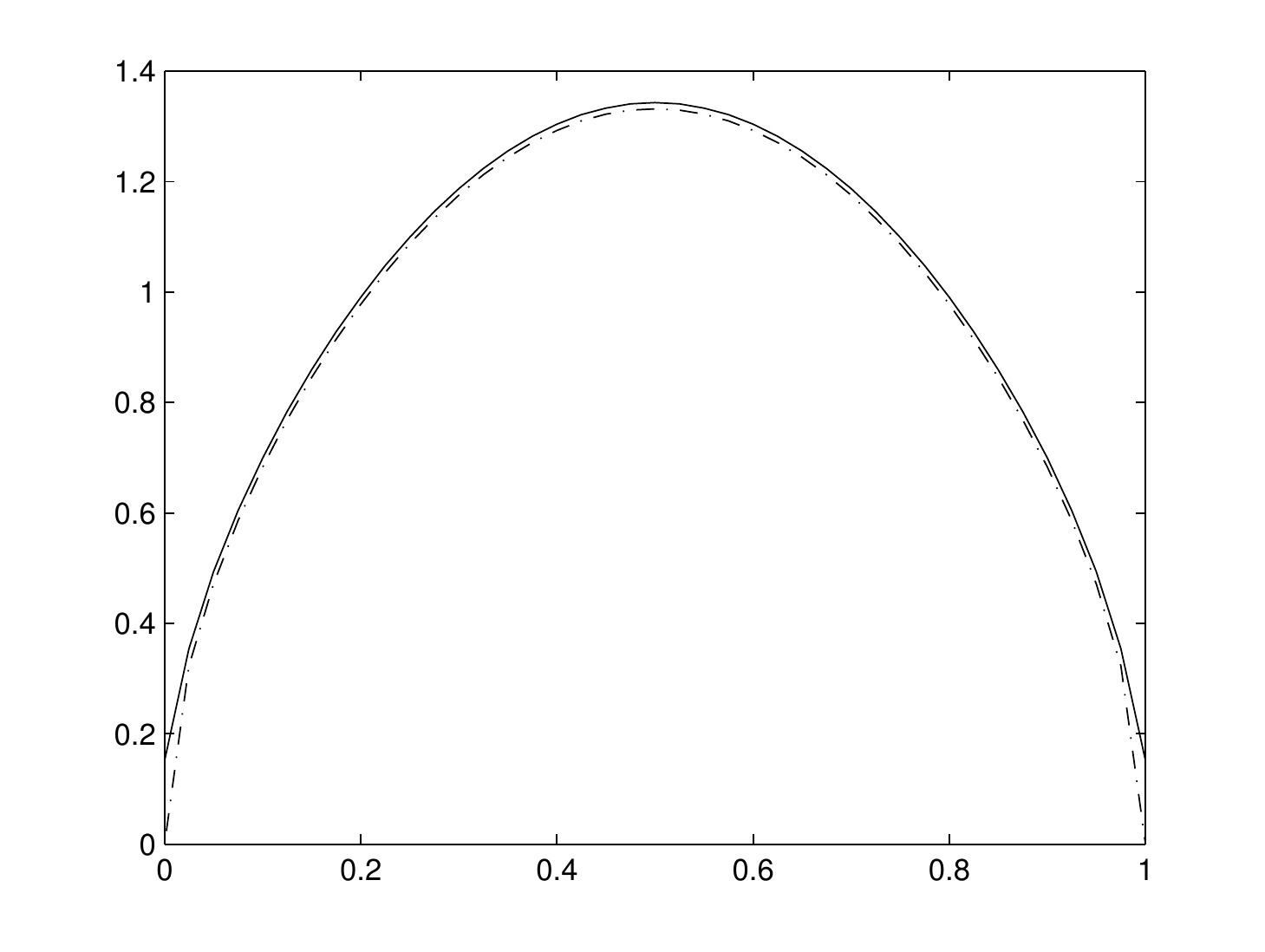}
\caption{Dependence on number of terms in Fourier series for $u_{0}$ at $y=a$. Solid line from $y\leq a$ expansion. Dashed line from $y\geq a$ expansion. (a) $N=2$, (b) $N=40$}
\label{fig:ES1}
\end{figure}

This physical interpretation then motivates a Fourier series approach
and using a rescaling of lengths and frequencies, $\hat{\xi}=\xi/2$,
$\hat{y}=y/2$, $\hat{a}=a/2$, $\hat{\Omega}=2\Omega$ and
$\hat{\kappa}=2\kappa$ gives the geometry investigated by
\cite{evans93a}. The full Rayleigh-Bloch solution for $u$ is obtained as
\begin{equation}
u(\hat{\xi},\hat{y})=\left\{    \begin{array}{ll}
\displaystyle{\sum_{n=0}^{\infty} A_{n}\cos p_{n}\hat{\xi} \cosh \alpha_{n}\hat{y}} & 0\le \hat{y}\leq\hat{a} \\[1em]
 \displaystyle{ \sum_{n=-\infty}^{\infty} B_{n}    {\rm e}^{{\rm i}\kappa_{n}\hat{\xi}} {\rm e}^{ -\gamma_{n}
(\hat{y}-\hat{a})}} 
& 
\hat{y}\geq\hat{a} 
\end{array}
\right. \label{eq:ES1}
\end{equation}
where $p_{n}=n\pi$, $\alpha_{n}=\sqrt{p_{n}^{2}-\hat{\Omega}^{2}}$, $\kappa_{n}=(2n\pi +\hat{\kappa})$, and $\gamma_{n}=\sqrt{\kappa_{n}^{2}-\hat{\Omega}^{2}}$. 
The coefficients are determined by imposing continuity of $u$ and
$\partial u/\partial\hat{y}$ at $\hat{y}=\hat{a}$, multiplying by
$\cos p_{m}\hat{\xi}$ and integrating across the cell width, which
allows the $A_{n}$ to be eliminated and leaves a set of linear
equations for the $\kappa_n B_{n}$ coefficients which are written in matrix notation as
\begin{equation}
\bf{M} \hspace{1ex}  (\boldmath{\kappa B}) = \bf{0} 
{\rm ,}\label{eq:ES2}
\end{equation}
 where ${\bf M}$ is a matrix that can be deduced from
 \cite{evans93a}. 
The dispersion relation is obtained by fixing values of $\hat{\kappa}$
and finding the corresponding values of $\hat{\Omega}$ for which
$\det ({\bf M})=0$, and then obtaining the eigensolutions for the
coefficients $\kappa_{n}B_{n}$. The standing wave eigensolution $u_0$
is the case where $\kappa =\pi/2$, $\hat{\kappa}=\pi$, $\Omega=\Omega_0$, and some of the
rows of $\bf{M}$ exhibit singularites. This then requires
modifications and the limiting value of the corresponding equations
must then be utilised in place of those of
\cite{evans93a}. Numerically the infinite summations are truncated for
some value of $N$ of modes and the infinite summations are replaced by
$\sum_{-N}^{N-1}$ and $\sum_{0}^{2N}$. For the standing waves  the $A_{n}$ are non-zero only for even values of $n$, and the $B_{n}$ satisfy $B_{n}+B_{-(n+1)}=0$.
As a consequence, $u_{0}$ is non-zero on the teeth of the comb for
$\hat{y}\leq\hat{a}$, and when repeated in the next strip with a sign
change exhibits a discontinuity at $\hat{y}=\hat{a}$. The Fourier
series converges to the mid-value, $0$, there, but the discontinuity
results in Gibb's phenomenon and requires a (fairly) large number of
terms, $N$, to be included in the summation to establish continuity of
$u_{0}$ and $\partial u_{0}/\partial \hat{y}$ for $0\leq\hat{\xi}\leq
1$ at $\hat{y}=\hat{a}$. 
The discontinuity in $u_{0}$ at $y=a$ if insufficient terms are included in the summation is illustrated in figure~\ref{fig:ES1}(a), which shows $N=2$. When $N=40$, shown in figure~\ref{fig:ES1}(b), there is good agreement. 

The expansion
\begin{equation}
u_{1}(\hat{\xi},\hat{y})=\left\{    \begin{array}{ll}
\displaystyle{-\sum_{n=0}^{\infty} A_{n}\hat{\xi}\cos p_{n}\hat{\xi} \cosh \alpha_{n}\hat{y}
+\sum_{n=0}^{\infty} X_{n}\cos p_{n}\hat{\xi} \cosh \alpha_{n}\hat{y}} & 0\le \hat{y}\leq\hat{a} \\[1em]
\displaystyle{{\rm i} \sum_{n=-\infty}^{\infty}\frac{ B_{n}\kappa_{n}}{\gamma_{n}}\hat{y}
{\rm e}^{{\rm i}\kappa_{n}\hat{\xi}} { \rm e}^{ -\gamma_{n}(\hat{y}-\hat{a})}
+\sum_{n=-\infty}^{\infty} Y_{n}
{\rm e}^{{\rm i}\kappa_{n}\hat{\xi}} { \rm e}^{ -\gamma_{n}(\hat{y}-\hat{a})}
} & \hat{y}\geq\hat{a} 
\end{array}
\right. \label{eq:ES3}
\end{equation}
satisfies the differential equation for $u_{1}$ (\ref{eq:firstOrder})
and the Bloch boundary conditions: For $U_1$ we again need to choose $\kappa=\pi/2$.
 The coefficients $X_{n}$ and $Y_{n}$ are to be determined by requiring continuity of $u_{1}$ and $\partial u_{1}/\partial\hat{y}$ at $\hat{y}=\hat{a}$. This leads to a matrix equation for the $Y_{n}$ coefficients
\begin{equation}
\bf{M} \hspace{1ex}  (\boldmath{\kappa Y}) = \bf{F} 
{\rm .}\label{eq:ES4}
\end{equation}
where $\bf{M}$ is the same (singular) matrix as in~(\ref{eq:ES2}) and $\bf{F}$ depends on the known coefficients $A_{n}$ and $B_{n}$. Hence, the solution for $u_{1}$ is arbitrary with respect to additional multiples of $u_{0}$, but these extra terms do not contribute to the coefficient $T$ and may be safely ignored. 
The integrals required to calculate $T$ are expressed in terms of the coefficients as:
\begin{equation}
\int\int_S u_0^2dS=\frac{A_{0}^{2}\hat{a}}{2}\left( 1+\frac{\sinh 2\alpha_{0}\hat{a}}{2\alpha_{0}\hat{a}}\right)
+\frac{\hat{a}}{4}\sum_{2}^{2N-1}A_{n}^{2}\left( 1+\frac{\sinh 2\alpha_{n}\hat{a}}{2\alpha_{n}\hat{a}}\right)
-\frac{1}{2}\sum_{-N}^{N-1}\frac{B_{n}^{2}}{\gamma_{n}}
{\rm ,}\label{eq:ES5}
\end{equation}
and
\begin{eqnarray}
\int\int_S  u_{1,\xi_1}u_0 -  u_{0,\xi_1}u_1 dS & = & 
-\frac{A_{0}^{2}\hat{a}}{2}\left( 1+\frac{\sinh 2\alpha_{0}\hat{a}}{2\alpha_{0}\hat{a}}\right)
-\frac{\hat{a}}{4}\sum_{2}^{2N-1}A_{n}^{2}\left( 1+\frac{\sinh 2\alpha_{n}\hat{a}}{2\alpha_{n}\hat{a}}\right)
\nonumber \\ & & \hspace{-10ex}
+\sum_{\stackrel{n=0}{ even}}^{2N-1}\!\!A_{n}\!\!
\sum_{\stackrel{m=1}{ odd}}^{2N-1}\!\!X_{m}
\left(\frac{n^{2}+m^{2}}{n^{2}-m^{2}}\right)
\left(\frac{\sinh (\alpha_{n}+\alpha_{m})\hat{a}}{\alpha_{n}+\alpha_{m}}
+\frac{\sinh (\alpha_{n}-\alpha_{m})\hat{a}}{\alpha_{n}-\alpha_{m}}
\right)
\nonumber \\ & & 
+\sum_{-N}^{N-1} \frac{B_{n}\kappa_{n}}{\gamma_{n}}
\left(-{\rm i}Y_{n} +\frac{B_{n}\kappa_{n}}{2\gamma_{n}^{2}}\left(2\gamma_{n}\hat{a}+1\right)\right)
{\rm .}\label{eq:ES6}
\end{eqnarray}

\begin{table}
\centering
\begin{tabular}{rlllll}\hline
$N$ &  2 & 5 & 10 & 20 & 40   \\
$\Omega_{0}$ &  0.2085 &  0.2101 &  0.2106 & 0.2108 &  0.2109 \\
$\Omega_{0}$ & 0.6244 &  0.6290 & 0.6305 &  0.6312 & 0.6315 \\
$\Omega_{0}$ &  1.0355 & 1.0431 & 1.0454 & 1.0466 & 1.0471 \\
$\Omega_{0}$ & 1.4307 & 1.4404 & 1.4433 & 1.4447 & 1.4454 \\ \hline
\end{tabular}
\caption{The four standing wave frequencies for the comb-like
  structure with $a=7$ calculated from Fourier series expansion with
  the convergence shown by increasing the modes used.}
\label{tab:ES1}
\end{table}

\begin{table}
\centering
\begin{tabular}{rlllll}\hline
$N$ &  2 & 5 & 10 & 20 & 40   \\
$T$ & -0.0068 & -0.0066 & -0.0066 & -0.0066 & -0.0066  \\
$T$ & -0.0704 & -0.0687  & -0.0686 & -0.0686 & -0.0686 \\
$T$ & -0.2877  & -0.2843 & -0.2849 & -0.2856 & -0.2860 \\
$T$ & -2.2022 & -2.3602 & -2.4267 & -2.4628 & -2.4816 \\ \hline
\end{tabular}
\caption{The values of $T$ associated with the four standing wave frequencies for the comb-like structure with $a=7$ calculated from Fourier series expansion with with
  the convergence shown by increasing the modes used.}
\label{tab:ES2}
\end{table}

The calculations of the standing wave frequencies and the coefficient $T$ are shown in tables ~\ref{tab:ES1} and \ref{tab:ES2} for different values of $N$ between 2 and 40 and demonstrate that these values are relatively insensitive to the value of $N$ used. There is good agreement with the values obtained from the full numerical simulation and the asymptotic approximation.

In Table \ref{tab:first_four} the  values of $T$ used in equation
(\ref{eq:effEqu}), which combined with the standing wave ($\Omega_0$)
and excitation ($\Omega$) frequencies yield an effective medium
equation. Fig. \ref{fig:seb2} shows the appearance of a new length
scale when a periodic comb-like structure with  $a=7$, is excited with
a line source at the frequencies of $\Omega=0.4509$ and
$\Omega=1.3138$ respectively in Fig. \ref{fig:seb2} (a) and (b). The
standing wave eigensolutions closest to these frequencies are shown in
Fig. \ref{fig:EigenComb}(a), (b) and show that on the microscale one
expects no oscillation or one oscillation along the open waveguide
formed by the comb teeth in one strip, and this local behaviour is
indeed seen in Fig. \ref{fig:seb2}. There is also clearly a long-scale
oscillation along the comb and the calculation of the apparent pseudo
wavelength is possible by HFH as explained in section
\ref{sec:general}\ref{sec:effMedia} and yields the respective
wavelengths $\lambda/2 \sim43.4$ and $\lambda/2\sim 73$. These are in accordance with panels (a) and (b) of Figs. \ref{fig:seb2} and \ref{fig:effComb} where the latter shows a complete reproduction by HFH of the numerical results, obtained by plotting Real$(u)$.    

\subsection{Array of cylinders}
\label{sec:cylinders}
Similarly to the comb structure one can also have a diffraction
grating constructed from a linear array of obstacles where surface
wave modes can again occur.
We consider  a linear periodic array of cylinders, as in say 
\cite{evans98}, where Rayleigh-Bloch modes are observed. 

The first
mode, which is symmetric about $y=0$, is shown in
Fig. \ref{fig:circleSymmetric}(b) and exists for all radii $r_0$ of
the cylinders such that
$r_0\in]0,1[$. Fig. \ref{fig:circleSymmetric}(a) shows the dispersion
branches for radii $r_0=0.4$, $0.5$ and $0.7$ and the associated HFH asymptotics. If the radius is
greater than $r_0\sim 0.81$ a second Rayleigh-Bloch mode appears,
illustrated in Fig. \ref{fig:circleAsymmetric}(b), which is
anti-symmetric about $y=0$. To motivate how this occurs we turn to a
two-dimensional rectangular lattice of cylinders, as a generalization
of \cite{antonakakis13a}, so instead of a grating we consider the
dispersion diagram of a doubly periodic structure where the width of
the rectangles is fixed to $2$ and the height is gradually increased
until a grating-like strip is
obtained. Fig. \ref{fig:Parametriccircle}(a), (b) and (c) show the
first three modes, and the light line $\Omega=\kappa$, for the
respective cell heights of $h=2$, $6$ and $30$ and each with a
centered hole of radius $r_0=0.95$. Both dispersion modes initially
above the light line converge to the latter as the height of the cell
increases and eventually one emerges beneath it. Upon inspection the
Bloch mode, for the rectangular array, that passes beneath the light
line has the appropriate symmetry and limits to the anti-symmetric
mode for the grating.  As discussed in \cite{evans98} the
critical radius value is $\sim 0.81$ and beyond this there is the
emergence of the antisymmetric trapped
mode; this is illustrated in Fig. \ref{fig:Parametriccircle}(d) which
shows the anti-symmetric mode, 
for a rectangular array height of $h=30$, for 
 radii $r_0=0.4$, $0.81$ and $0.95$ respectively. 
For radii $r_0=0.4$ and $0.81$ the mode merges with the light line,
but the mode related to $0.95$ emerges below the light line and one
then observes this anti-symmetric Rayleigh Bloch mode. For all radii
less than $\sim 0.81$ all modes, bar the first, will collapse on the
lightline. Fig. \ref{fig:Parametriccircle}(e) shows a summary of the
variation of the standing wave frequencies with radius, and the
appearance of this anti-symmetric mode for radii in the interval
$[0.81,1[$ is evident. Notably, the asymptotic HFH theory captures the
behaviour of the dispersion curves in this case too as shown in
Fig. \ref{fig:circleAsymmetric}(a). 

 To illustrate further HFH and the emergence of the long-scale
 oscillation we performed large-scale finite element simulations
 summarised in  Fig. \ref{fig:seb1}. The asymmetric mode was generated
 using a dipole source, to trigger the asymmetry, and the symmetric
 mode using a line source. The chosen frequencies 
 are slightly away from the standing wave frequencies and are
 respectively $\Omega=1.508$ and $\Omega=1.38$ for panels (a) and
 (b). Once again the apparent lengthscales are evaluated by HFH to be
 $\lambda/2=109.4$ and $\lambda/2=72.7$ which are confirmed by the
 numerics as well as in Fig. \ref{fig:effCircle}.

\begin{figure}
\centering
\includegraphics[scale=0.6]{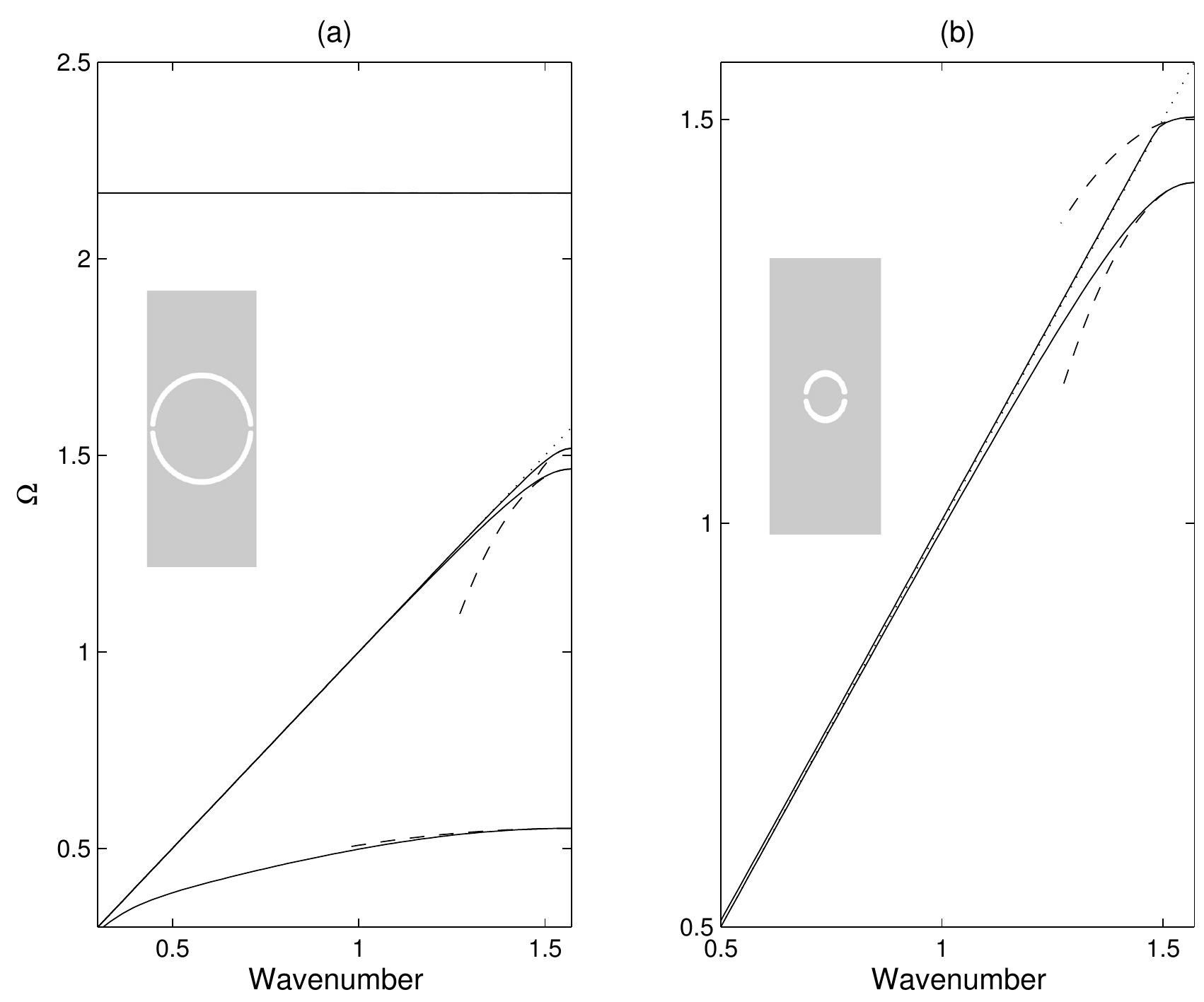}
 \caption{The dispersion branches for the SRR structure. Results from
   numerical simulations shown as solid lines, the asymptotics as
   dashed and the light line is dotted. Panel (a) is for a large SRR
   of outer radius $R_{out}=0.95$ and inner radius $R_{in}=0.85$ and
   panel (b) is for a smaller SRR with $R_{out}=0.4$ and
   $R_{in}=0.3$.}
\label{fig:SRR_Dispersion}
\end{figure}

\begin{table}
\centering
\begin{tabular}{cccc}\hline
$T$ & $ \Omega_0$ \\
$ -0.145181377783699$ & $ 0.550884858382472$ \\
$ -12.037628085319582$ & $ 1.465518146041600$ \\
$ -25.144195967619925$ & $ 1.517732423423271$ \\
$ -0.003773469173013$ & $ 2.167224509645187$ \\
\hline
\end{tabular}
\caption{The four standing wave frequencies, below the cutoff, for a SRR grating with $R_{out}=0.95$ and $R_{in}=0.85$,
  c.f. Fig. \ref{fig:SRR_Dispersion}(a), together with associated values for $T$.}
\label{tab:SRR}
\end{table}

\begin{figure}
\centering
\includegraphics[scale=0.5]{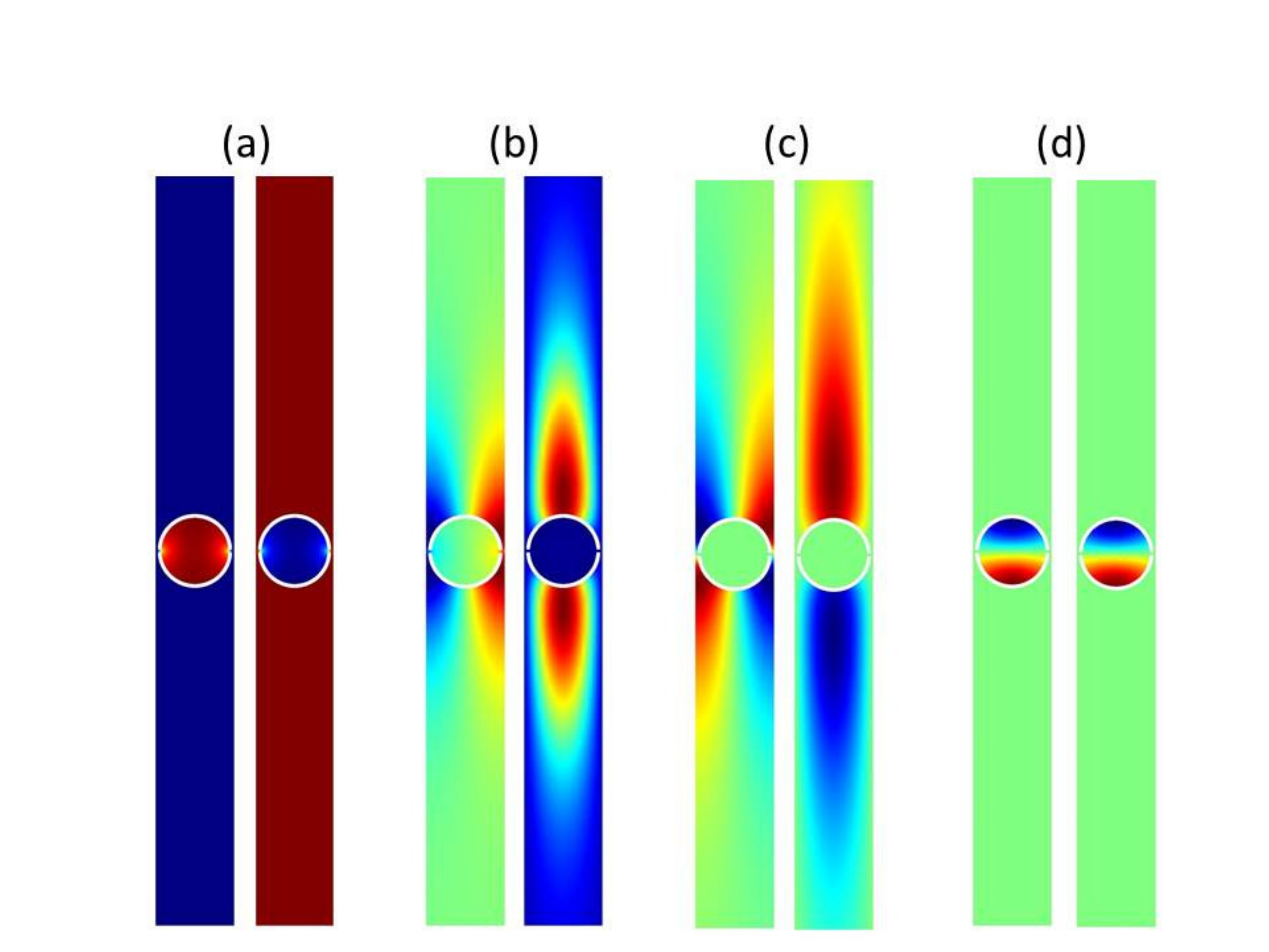}
 \caption{The eigenfunctions $U_0$, and $U_1$ shown for the
   SRR structure, with outer, inner radius $R_{out}=0.95$, $R_{in}=0.85$ c.f. Fig. \ref{fig:SRR_Dispersion}(a). These are for the $\Omega_0$ in Table \ref{tab:SRR} with (a)-(d)
   for ascending $\Omega_0$. In each panel $U_0$ is shown on the left
   and $U_1$ on the right.}
\label{fig:SRR_Eigenvectors}
\end{figure}

\subsection{Array of Split Ring Resonators}
\label{sec:SRR}
SRR are extensively used to achieve left handed materials used in
electromagnetism \cite[]{ramakrishna05a}. For SRRs here we choose to use a
simple cylindrical annulus with two ligaments connecting the inner
cylinder to the outer material. The weak coupling between the inner
cylinder through these two thin ligaments is important as this
arrangement can  act as local
resonators and this micro-resonance is important in photonic
applications and in metamaterials \cite{pendry99a}. 
In
SRR gratings Rayleigh-Bloch modes occur at frequencies above the
cutoff due to this resonance behaviour within the inner part
of the SRR as shown in the fourth mode of
Figs. \ref{fig:SRR_Dispersion}(a) and the resonance is clear in the
eigensolution shown in
\ref{fig:SRR_Eigenvectors}(d).  The ultra-flat dispersion curve,
Fig. \ref{fig:SRR_Dispersion}(a), is associated with dipole localized
modes in every SRR of the grating and it can be predicted using a
geometric asymptotic technique discussed in
\cite{antonakakis13a,movchan04a}. 

The modes that arise for the grating of
SRR split into two families, one which is very similar to those of the cylinders of
the last section, that is, 
Fig. \ref{fig:SRR_Eigenvectors}(b) and (c) are respectively similar to
those of Figs. \ref{fig:circleSymmetric}(b) and
\ref{fig:circleAsymmetric}(b). The lowest mode, whose eigensolution is
shown in Fig. \ref{fig:SRR_Eigenvectors}(a), is again one primarily
associated with the inner cylinder and vibrations of the ligaments. 

HFH is used to generate the asymptotics and 
 Table \ref{tab:SRR} shows the standing
wave frequencies and respective values of $T$ for the first four modes
of a SRR grating with outer radius of $R_{out}=0.95$. The asymptotics
of the dispersion curves again show pleasing accuracy. 

\begin{figure}
\centering
\includegraphics[scale=0.17]{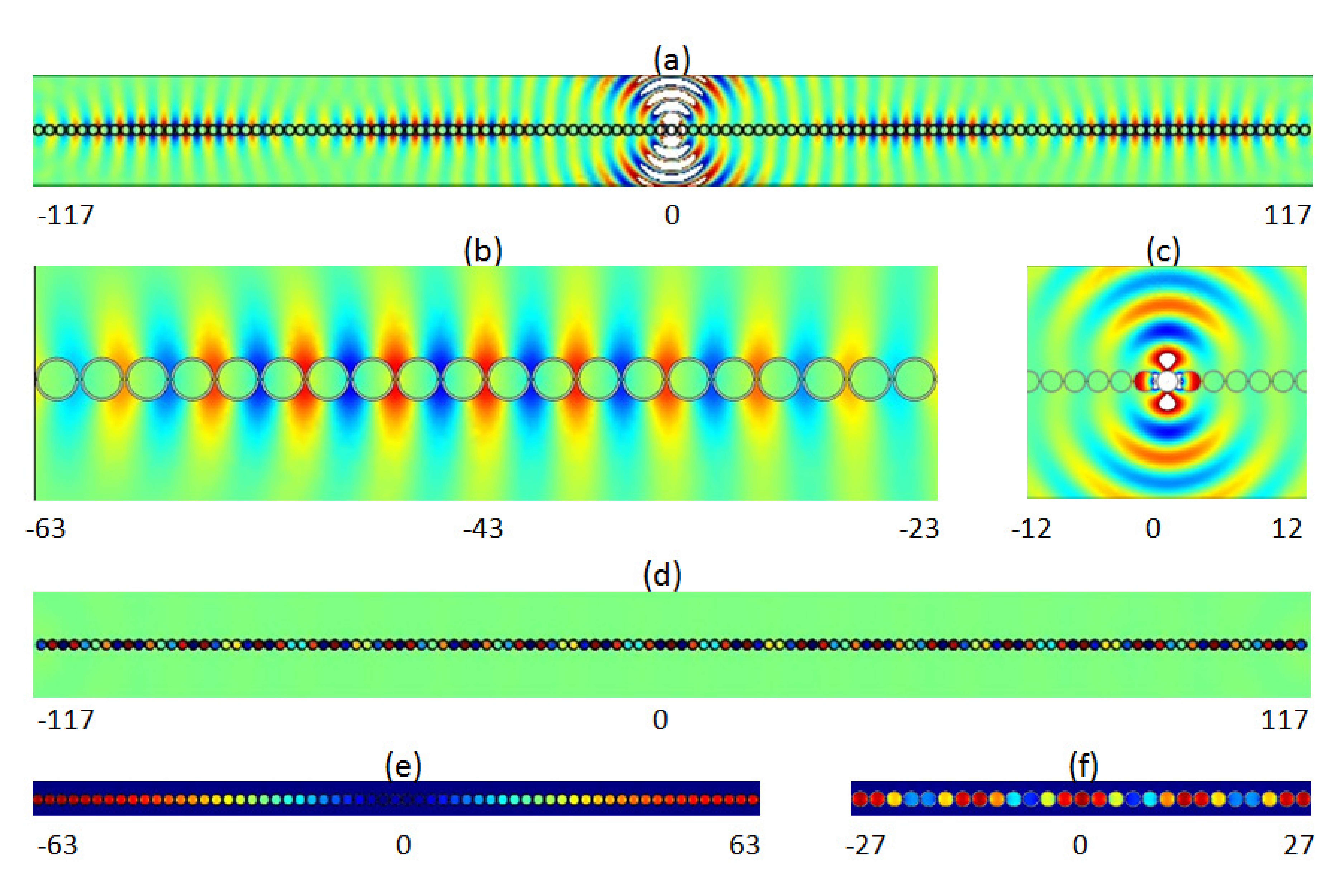}
 \caption{Plots of real $(u)$ for a SRR grating
with SRR of inner and outer radii $0.85$ and $0.95$ respectively and 
ligaments of thickness $0.06$ (standing wave frequencies in
Table \ref{tab:SRR}): (a) Field generated by a line source at
$\Omega=1.45$; (b,c) Detail of real $(u)$ at $\Omega=1.45$ centered
 around $x=-43$ showing the developed field (b) and
around the source (c); (d) Field generated by a line source at $\Omega=0.54$; (e,f) Close up on the
absolute value of field $u$ at frequency $\Omega=0.55, \Omega=0.54$ respectively.}
\label{fig:SRR_seb}
\end{figure}

\begin{figure}
\centering
\includegraphics[scale=0.17]{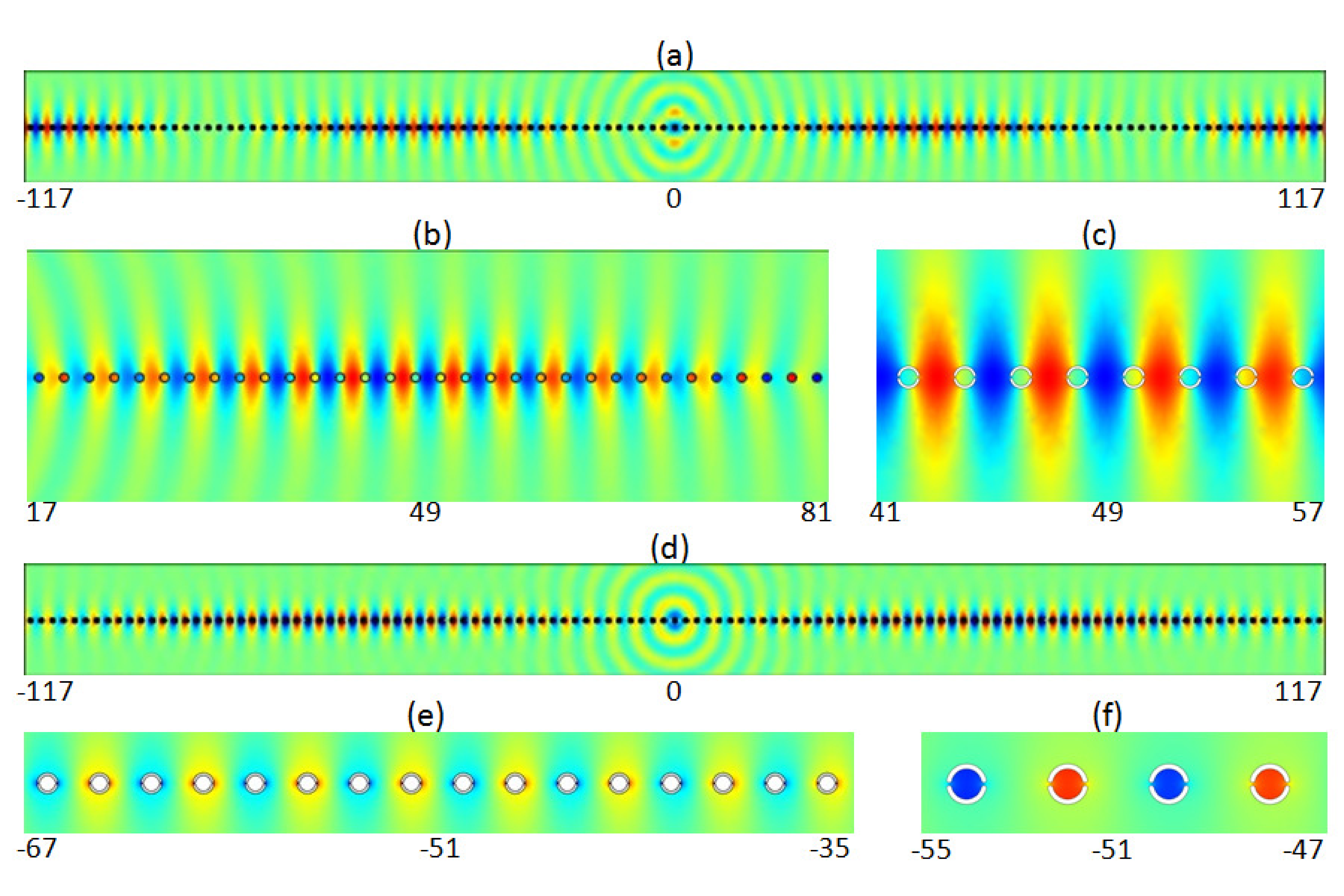}
 \caption{Plots of real $(u)$ for a SRR grating
with SRR of inner and outer radii $0.3$ and $0.4$ and ligamets of
thickness $0.06$: (a) Fields generated by a line source at
$\Omega=1.50$ ($\Omega_0=1.50295$); (b,c) Detail of $u$ at
$\Omega=1.50$ centered
 around $x=49$ showing the developed field
; (d)
Field generated by a line source at $\Omega=1.42$
($\Omega_0=1.42199$); (e,f) Detail of real $(u)$ at frequency $\Omega=1.42$}
\label{fig:SRR_seb4}
\end{figure}

\begin{figure}
\centering
\includegraphics[scale=0.57]{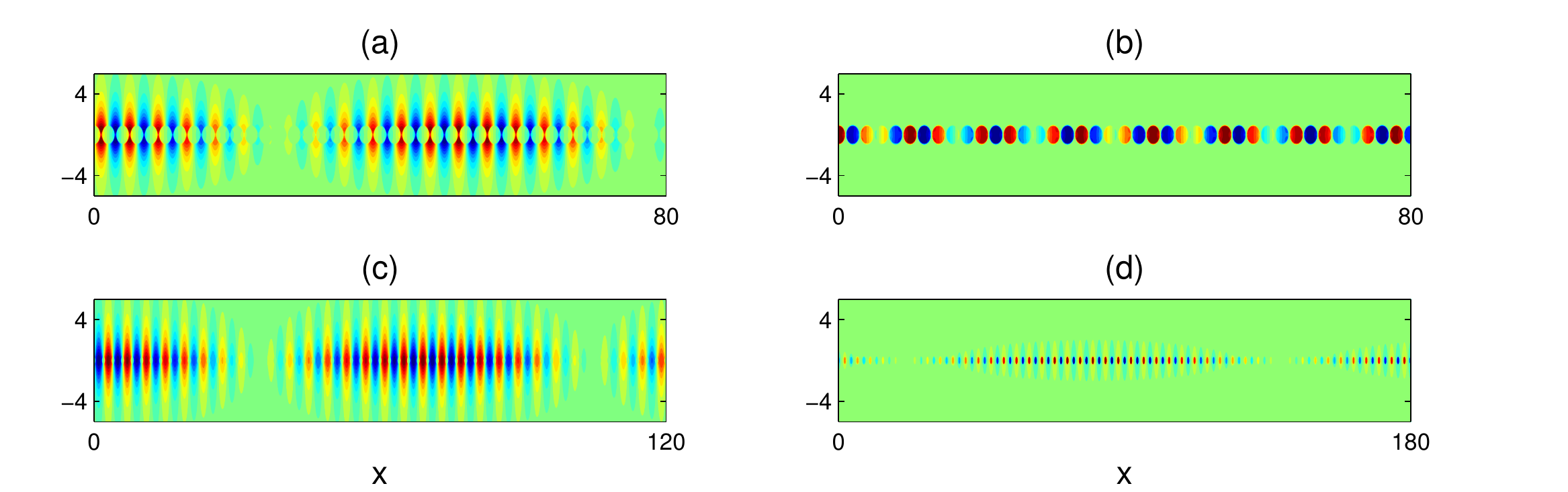}
 \caption{Panels (a) and (b) show real $(u)$ from HFH for a SRR, 
with inner and outer radii of $R_{in}=0.85$ and $R_{out}=0.95$, c.f. Fig. \ref{fig:SRR_seb}, 
generated by a line source at $\Omega=1.45$ and $\Omega=0.54$
respectively. Panels (c) and (d) show real $(u)$ from HFH for a SRR, 
with inner and outer radii of $R_{in}=0.3$ and $R_{out}=0.4$, c.f. Fig. \ref{fig:SRR_seb4}, 
generated by a line source at $\Omega=1.50$ and $\Omega=1.42$ respectively.}
\label{fig:effSRR}
\end{figure}

Numerical finite element solutions for line source excitation show
plainly this separation into exterior modes akin to those of the
cylinder  Figs. \ref{fig:SRR_seb}(a,b,c) and those localised almost
entirely within the SRR as in Figs \ref{fig:SRR_seb}(d,e,f): In these
latter cases the array acts very clearly as an oscillating string. The
smaller SRR illustrated in Fig \ref{fig:SRR_seb4} gives an even more
pronounced locally anti-periodic oscillation with long-scale
oscillation. 
The excitation frequencies are chosen to be close to those of standing waves and
the long-scale behaviour extracted using HFH as seen in the Fig. \ref{fig:effSRR}. The wavelengths associated with Fig. \ref{fig:effSRR} are in the panel's order of appearance, $\lambda/2=51.2,11,69.6,119.1$.

\section{Defect states in quasi periodic gratings}
\label{sec:defectState}
The previous examples illustrate HFH for perfectly periodic
media but its applications go further than this. In section
\ref{sec:examples} HFH asymptotics and the resulting effective media successfully
homogenise perfect periodic arrays, but one could also obtain
analytical or numerical solutions fairly quickly at least for simple
geometries. The real power of HFH lies in its capability to move away
from perfect periodicity, we now take the comb of section
\ref{sec:examples}\ref{sec:comb}, but now vary the height of the
comb's teeth with
respect to the $x$ coordinate by a function $g(X)$, so that their
height is
\(
 a(1-\epsilon^2 g(X))
\)
 and ask the
question of whether localized states exist at specific frequencies in such quasi-periodic media,
that is, are there finite energy states that have exponential decay
along the array? 
We make the following change of coordinates in order to transform the
varying tooth height in $x_2$ to constant height pins in the new coordinate $\xi_2$ such that,
\beq
\xi_1=\frac{x_1}{l}\text{,}\quad X=\frac{x_1}{L}\text{,}\quad\xi_2=\frac{x_2}{l}\left[1-\epsilon^2g(X)\right].
\label{eq:variables}
\eeq
This sleight of hand transforms the medium and moves the tooth heights
to a constant within this transformed medium. Following through the
asymptotic procedure,  
as in section \ref{sec:general}, we obtain three equations ordered in
$\epsilon$, the only change is at 
 second order where (\ref{eq:secondOrder})  becomes 
\beq
  u_{2,\xi_i\xi_i} + \Omega_0^2 u_2 =-u_{0,XX}
   -2u_{1,\xi_1 X} +2g(X)u_{0,\xi_2\xi_2}
-\Omega_1^2 u_1 -\Omega_2^2 u_0 
  \label{eq:secondOrderDefect}
\eeq
which contains an additional term.
Neumann boundary conditions remain unchanged for leading and first order but in second order yield,
\beq
u_{1,X}n_1 + u_{2,\xi_i} n_i-g(X)u_{0,\xi_2}n_2=0.
\label{BCDefect}
\eeq
Using a solvability condition we obtain an equation for $f_0$ as,
\beq
Tf_{0,XX}+f_0[\alpha g(X)+\Omega_2^2]=0,\quad\text{with}\quad \alpha=\frac{\dint_S(U_{0,\xi_2}^2-U_0U_{0,\xi_2\xi_2})dS}{\dint_SU_0^2dS},
\label{eq:f0Defect}
\eeq 
where $T$ is given in (\ref{eq:t11simple}). This is a 
 Schroedinger equation and for specific choices of $g(X)$ exact
solutions exist notably for $g(X)=-{\rm sech}^2X$ as in
\cite[]{infeld51a,craster10b}, hence adopting this variation an asymptotic value of the lowest
defect mode frequency is explicitly 
\beq
    \Omega^2=\Omega_0^2-\frac{T\epsilon^2}{4}\left(1-
      \sqrt{1-4\alpha/T}\right)^2 
\label{eq:asy}
\eeq
provided that $\alpha/T$ is always negative, which occurs as $T$ is always negative and $\alpha$ positive. The associated solutions for $f_0(X)$ are \cite[]{infeld51a},
\beq
f_0(X)=\pi^{-\frac{1}{4}}\bigg(\frac{\Gamma(\gamma)}{\Gamma(\gamma-1/2)}\bigg)^{\frac{1}{2}}{\rm cosh}^{-\gamma+1/2}X,
\label{eq:f0Localized}
\eeq
where for the lowest defect mode $\gamma=\sqrt{1/4-\alpha/T}$ and
$\Gamma(\gamma)$ is the  Gamma function
\cite{abramowitz64a}.
 
\begin{figure}
\centering
\includegraphics[scale=0.8]{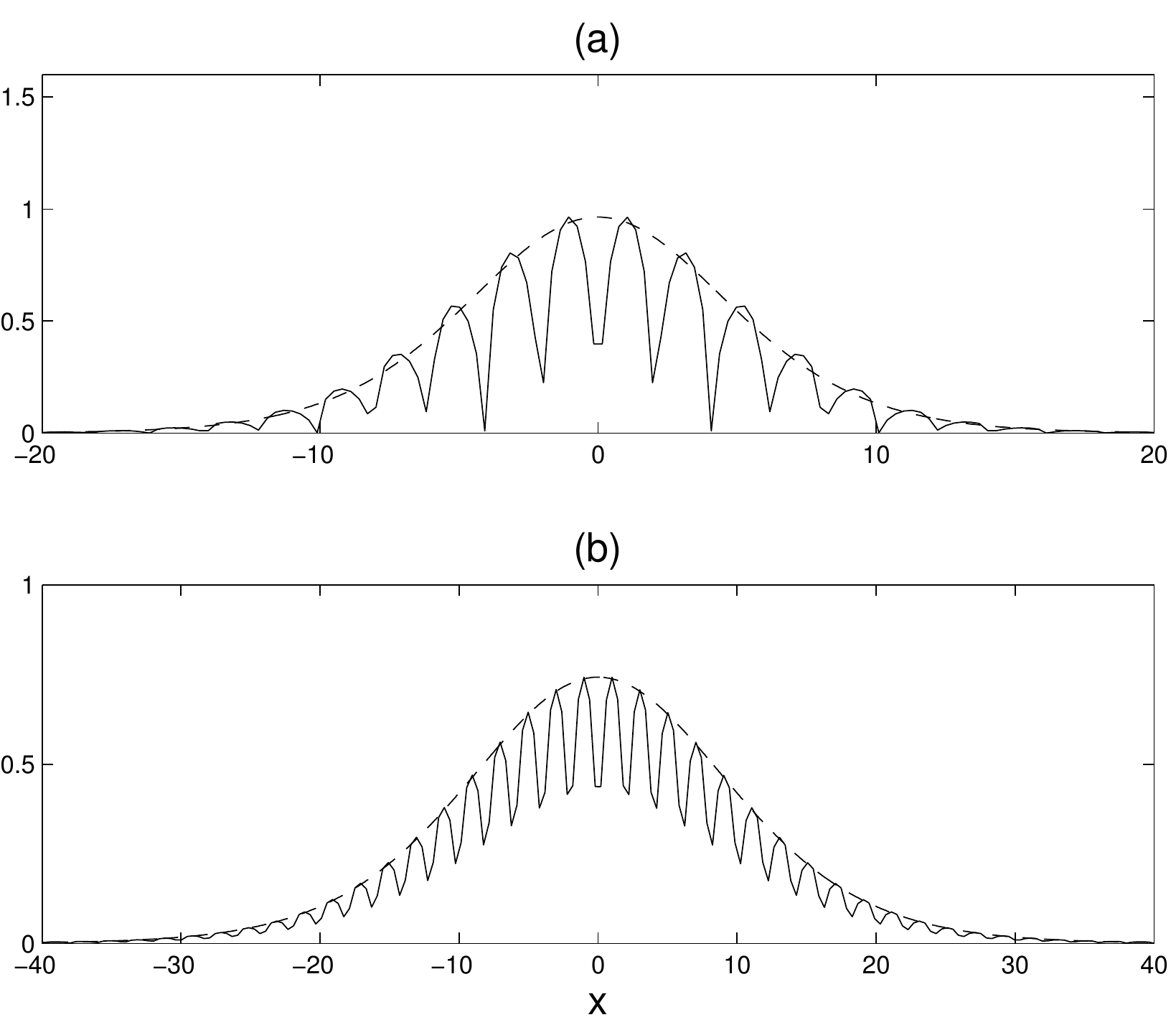}
 \caption{The localized defect mode shown for the comb-like structure
   withvariation of tooth height for $a=7$. The variation follows equation (\ref{eq:variables}) with $\epsilon=0.125$ in panel (a) and $\epsilon=0.0625$ in panel (b). In solid are solutions from FEM simulations of $u$ along $x$ and for $y=7$, and in dashed are from equation (\ref{eq:f0Localized}).}
\label{fig:local}
\end{figure}

For $a=7$ Table \ref{tab:local} shows the predictions of the
frequencies at which these defect states arise versus values extracted
from finite element simulations which are reassuringly
accurate, and these defect mode frequencies are above the standing
wave frequencies as one would expect. 
Perhaps more compelling are the 
 illustrative solutions shown in Fig.  \ref{fig:local} which show
 $f_0$ versus the numerical eigensolutions; as both solutions are
 arbitrary to within a multiplicative constant we normalise to have
 $\max(f_0)$ equal to the maximum value from the numerics.

\begin{table}
\centering
\begin{tabular}{ccc}\hline
$\epsilon$ & $ \Omega_{HFH}$ & $\Omega_{num}$\\
$ 0.125$ & $0.21247 $ & $0.21252$\\
$ 0.0625$ & $0.21074 $ & $0.21079$\\
\hline
\end{tabular}
\caption{The predicted frequencies of the localized defect mode near
  the first standing wave frequency ($\Omega_0=0.210161050669707$,
  c.f. Table \ref{tab:first_four}) for the comb-like structure with
  $a=7$ c.f. Fig. \ref{fig:comb}(d). The frequencies $\Omega_{HFH}$
  come from the asymptotics (\ref{eq:asy}) whereas $\Omega_{num}$
  gives predictions from FEM simulations. The parameter $\epsilon$
  controls the variation of tooth height in (\ref{eq:variables}).}
\label{tab:local}
\end{table}

\section{Concluding remarks}
\label{sec:conclude}
It is shown here that one can take a microstructured surface, or
diffraction grating \cite[]{popov12a,petit80a}, and close to the standing wave frequencies that
occur, one can represent the surface as an effective string, or
membrane. The standing waves can occur at high frequencies and as a
result  the effective stiffness is not simply an average but involves
the integrals over a microscale, importantly the effective equation is
posed entirely on the long-scale with the short-scale built in through
integrated quantities. Thus we extend homogenization in two distinct
directions enabling microstructed surfaces, instead of the more usual
bulk media, to be modelled and away from the usual low-frequency
limit. Given the effective equation description one can then concentrate
 numerical efforts on modeling instead of capturing the fine scale detail.
Indeed, as shown in section \ref{sec:defectState}, one can use the
effective description to capture analytically features such as defect
states caused by non-periodic behaviour. 

There are several practical directions that could be pursued using
this analysis, notably the surface wave for line source excitation
demonstrates the two-scale behaviour beautifully with a short-scale
oscillation from one neighbouring strip to the next and, in some sense,
chooses its own longer wavelength. The current theory neatly
encapsulates this and this information could be used as part of an
inverse problem to determine the quality of microscale or nanoscale
surfaces, and the defect states could identify local
damage. Importantly, questions related to tuning a surface to have
designer properties can be encapsulated into how the coefficient $T$
behaves and that too avoids lengthy computations using numerical methods for gratings such as Fourier \cite[]{li96a} or differential \cite[]{lalanne97a} methods.

In summary, one can now take a microstructured surface, or diffraction
grating, that is periodic, or nearly so, and replace it by a continuum
description that captures the surface Rayleigh-Bloch waves.

\section*{Acknowledgements}
 RVC and EAS thank the EPSRC (UK) 
for their support through research grants EP/I018948/1 \&  EP/J009636/1. 
 SG is thankful for an ERC starting grant (ANAMORPHISM) which facilitates the collaboration with Imperial College London.

\bibliographystyle{jfm}


\begin{thebibliography}{50}
\expandafter\ifx\csname natexlab\endcsname\relax\def\natexlab#1{#1}\fi

\bibitem[Abramowitz \& Stegun(1964)]{abramowitz64a}
{\sc Abramowitz, M. \& Stegun, I.~A.} 1964 {\em Handbook of Mathematical
  Functions\/}. Washington: National Bureau of Standards.

\bibitem[Allaire \& Piatnitski(2005)]{allaire05a}
{\sc Allaire, G. \& Piatnitski, A.} 2005 Homogenisation of the {S}chr\"odinger
  equation and effective mass theorems. {\em Commun. Math. Phys.\/} {\bf 258},
  1--22.

\bibitem[Antonakakis \& Craster(2012)]{antonakakis12a}
{\sc Antonakakis, T. \& Craster, R.~V.} 2012 High frequency asymptotics for
  microstructured thin elastic plates and platonics. {\em Proc. R. Soc. Lond.
  {\rm A}\/} {\bf 468}, 1408--1427.

\bibitem[Antonakakis {\em et~al.\/}(2013)Antonakakis, Craster \&
  Guenneau]{antonakakis13a}
{\sc Antonakakis, T., Craster, R.~V. \& Guenneau, S.} 2013 Asymptotics for
  metamaterials and photonic crystals. {\em Proc. R. Soc. Lond. {\rm A}\/} {\bf
  469}, 20120533.

\bibitem[Bakhvalov \& Panasenko(1989)]{bakhvalov89a}
{\sc Bakhvalov, N. \& Panasenko, G.} 1989 {\em Homogenization: Averaging
  Processes in Periodic Media\/}. Amsterdam: Kluwer.

\bibitem[Barlow \& Karbowiak(1954)]{barlow54a}
{\sc Barlow, H. E.~M. \& Karbowiak, A.~E.} 1954 An experimental investigation
  of the properties of corrugated cylindrical surface waveguides. {\em Proc.
  IEE\/} {\bf 101}, 182--188.

\bibitem[Bensoussan {\em et~al.\/}(1978)Bensoussan, Lions \&
  Papanicolaou]{bensoussan78a}
{\sc Bensoussan, A., Lions, J. \& Papanicolaou, G.} 1978 {\em Asymptotic
  analysis for periodic structures\/}. North-Holland, Amsterdam.

\bibitem[Birman \& Suslina(2006)]{birman06a}
{\sc Birman, M.~S. \& Suslina, T.~A.} 2006 Homogenization of a multidimensional
  periodic elliptic operator in a neighborhood of the edge of an internal gap.
  {\em Journal of Mathematical Sciences\/} {\bf 136}, 3682--3690.

\bibitem[{Bonnet-Bendhia} \& Starling(1994)]{bonnet94a}
{\sc {Bonnet-Bendhia}, A.~S. \& Starling, F.} 1994 Guided waves by
  electromagnetic gratings and nonuniqueness examples for the diffraction
  problem. {\em Math. Meth. Appl. Sci.\/} {\bf 17}, 305--338.

\bibitem[Brekhovskikh(1959)]{brekhovskikh59a}
{\sc Brekhovskikh, L.~M.} 1959 Surface waves in acoustics. {\em Sov. Phys.
  Acoust.\/} {\bf 5}, 3--12.

\bibitem[Conca {\em et~al.\/}(1995)Conca, Planchard \& Vanninathan]{conca95a}
{\sc Conca, C., Planchard, J. \& Vanninathan, M.} 1995 {\em Fluids and Periodic
  structures\/}. Res. Appl. Math., Masson, Paris.

\bibitem[Craster {\em et~al.\/}(2013)Craster, Joseph \& Kaplunov]{craster13a}
{\sc Craster, R.~V., Joseph, L.~M. \& Kaplunov, J.} 2013 Long-wave asymptotic
  theories: {T}he connection between functionally graded waveguides and
  periodic media. Under review.

\bibitem[Craster {\em et~al.\/}(2011)Craster, Kaplunov, Nolde \&
  Guenneau]{craster11a}
{\sc Craster, R.~V., Kaplunov, J., Nolde, E. \& Guenneau, S.} 2011 High
  frequency homogenization for checkerboard structures: Defect modes,
  ultra-refraction and all-angle-negative refraction. {\em J. Opt. Soc. Amer.
  A\/} {\bf 28}, 1032--1041.

\bibitem[Craster {\em et~al.\/}(2010{\natexlab{{\em a\/}}})Craster, Kaplunov \&
  Pichugin]{craster10a}
{\sc Craster, R.~V., Kaplunov, J. \& Pichugin, A.~V.} 2010{\natexlab{{\em
  a\/}}} High frequency homogenization for periodic media. {\em Proc R Soc Lond
  A\/} {\bf 466}, 2341--2362.

\bibitem[Craster {\em et~al.\/}(2010{\natexlab{{\em b\/}}})Craster, Kaplunov \&
  Postnova]{craster10b}
{\sc Craster, R.~V., Kaplunov, J. \& Postnova, J.} 2010{\natexlab{{\em b\/}}}
  High frequency asymptotics, homogenization and localization for lattices.
  {\em Q. Jl. Mech. Appl. Math.\/} {\bf 63}, 497--519.

\bibitem[De{S}anto(1972)]{desanto72a}
{\sc De{S}anto, J.~A.} 1972 Scattering from a periodic corrugated structure
  {II}. {T}hin comb with hard boundaries. {\em J. Math. Phys.\/} {\bf 13},
  336--341.

\bibitem[Enoch \& Bonod(2012)]{enoch12a}
{\sc Enoch, S. \& Bonod, N.} 2012 {\em Plasmonics: From Basics to Advanced
  Topics\/}. Springer Series in Optical Sciences, Vol. 167.

\bibitem[Evans \& Linton(1993)]{evans93a}
{\sc Evans, D.~V. \& Linton, C.~M.} 1993 Edge waves along periodic coastlines.
  {\em QJMAM\/} {\bf 46}, 643--656.

\bibitem[Evans \& Porter(1998)]{evans98}
{\sc Evans, D.~V. \& Porter, R.} 1998 Trapping and near-trapping by arrays of
  cylinders in waves. {\em J. Engng Math.\/} {\bf 35}, 149--179.

\bibitem[Evans \& Porter(2002)]{evans02a}
{\sc Evans, D.~V. \& Porter, R.} 2002 On the existence of embedded surface
  waves along arrays of parallel plates. {\em QJMAM\/} {\bf 55}, 481--494.

\bibitem[{Fernandez-Dominguez} {\em et~al.\/}(2011){Fernandez-Dominguez},
  {Garcia-Vidal} \& {Martin-Moreno}]{fernandez11a}
{\sc {Fernandez-Dominguez}, A.~I., {Garcia-Vidal}, F. \& {Martin-Moreno}, L.}
  2011 {\em Structured surfaces as optical metamaterials, Ed. A. A.
  Maradudin\/}, chap. Surface electromagnetic waves on structured perfectly
  conducting surfaces. CUP.

\bibitem[Gridin {\em et~al.\/}(2004)Gridin, Adamou \& Craster]{gridin04a}
{\sc Gridin, D., Adamou, A. T.~I. \& Craster, R.~V.} 2004 Electronic
  eigenstates in quantum rings: Asymptotics and numerics. {\em Phys Rev B\/}
  {\bf 69}, 155317.

\bibitem[Gridin {\em et~al.\/}(2005)Gridin, Craster \& Adamou]{gridin05a}
{\sc Gridin, D., Craster, R.~V. \& Adamou, A. T.~I.} 2005 Trapped modes in
  curved elastic plates. {\em Proc R Soc Lond A\/} {\bf 461}, 1181--1197.

\bibitem[Hoefer \& Weinstein(2011)]{hoefer11a}
{\sc Hoefer, M.~A. \& Weinstein, M.~I.} 2011 Defect modes and homogenization of
  periodic {S}chr\"odinger operators. {\em SIAM J. Math. Anal.\/} {\bf 43},
  971--996.

\bibitem[Hurd(1954)]{hurd54a}
{\sc Hurd, R.~A.} 1954 The propagation of an electromagnetic wave along an
  infinite corrugated surface. {\em Can. J. Phys.\/} {\bf 32}, 727--734.

\bibitem[Infeld \& Hull(1951)]{infeld51a}
{\sc Infeld, L. \& Hull, T.~E.} 1951 The factorization method. {\em Rev. Modern
  Phys.\/} {\bf 23}, 21--68.

\bibitem[Joannopoulos {\em et~al.\/}(2008)Joannopoulos, Johnson, Winn \&
  Meade]{joannopoulos08a}
{\sc Joannopoulos, J.~D., Johnson, S.~G., Winn, J.~N. \& Meade, R.~D.} 2008
  {\em Photonic Crystals, Molding the Flow of Light\/}, 2nd edn. Princeton
  University Press, Princeton.

\bibitem[Joseph \& Craster(2013)]{joseph13a}
{\sc Joseph, L.~M. \& Craster, R.~V.} 2013 Asymptotics for {Rayleigh-Bloch}
  waves along lattice line defects. Multiscale Modeling and Simulation.

\bibitem[Kaplunov {\em et~al.\/}(2005)Kaplunov, Rogerson \&
  Tovstik]{kaplunov05a}
{\sc Kaplunov, J.~D., Rogerson, G.~A. \& Tovstik, P.~E.} 2005 Localized
  vibration in elastic structures with slowly varying thickness. {\em Quart. J.
  Mech. Appl. Math.\/} {\bf 58}, 645--664.

\bibitem[Lalanne(1997)]{lalanne97a}
{\sc Lalanne, P.} 1997 Convergence performance of the coupled-wave and the
  differential method for thin gratings. {\em J. Opt. Soc. Am. A\/} {\bf 14},
  1583--1591.

\bibitem[Li(1996)]{li96a}
{\sc Li, L.} 1996 Use of {F}ourier series in the analysis of discontinuous
  periodic structures. {\em J. Opt. Soc. Am. A\/} {\bf 13}, 1870--1876.

\bibitem[Linton \& Mc{I}ver(2002)]{linton02a}
{\sc Linton, C.~M. \& Mc{I}ver, M.} 2002 The existence of {R}ayleig{h-B}loch
  surface waves. {\em J. Fluid Mech.\/} {\bf 470}, 85--90.

\bibitem[Maier(2007)]{maier07a}
{\sc Maier, S.~A.} 2007 {\em Plasmonics: Fundamentals and Applications\/}.
  Springer-Verlag.

\bibitem[Makwana \& Craster(2012)]{makwana13a}
{\sc Makwana, M. \& Craster, R.~V.} 2012 Localized defect states for high
  frequency homogenized lattice models. {\em Quart. J. Mech. Appl. Math.\/} To
  appear: doi:10.1093/qjmam/hbt005.

\bibitem[Maniar \& Newman(1997)]{maniar97a}
{\sc Maniar, H.~D. \& Newman, J.~N.} 1997 Wave diffraction by a long array of
  cylinders. {\em J. Fluid Mech.\/} {\bf 339}, 309–330.

\bibitem[Mc{I}ver {\em et~al.\/}(1998)Mc{I}ver, Linton \& Mc{I}ver]{mciver98a}
{\sc Mc{I}ver, P., Linton, C.~M. \& Mc{I}ver, M.} 1998 Construction of trapped
  modes for wave guides and diffraction gratings. {\em Proc. R. Soc. Lond. {\rm
  A}\/} {\bf 454}, 2593--2616.

\bibitem[Movchan \& Guenneau(2004)]{movchan04a}
{\sc Movchan, A.~B. \& Guenneau, S.} 2004 Split-ring resonators and localized
  modes. {\em Phys. Rev. B\/} {\bf 70}, 125116.

\bibitem[Nemat-Nasser {\em et~al.\/}(2011)Nemat-Nasser, Willis, Srivastava \&
  Amirkhizi]{nematnasser11a}
{\sc Nemat-Nasser, S., Willis, J.~R., Srivastava, A. \& Amirkhizi, A.~V.} 2011
  Homogenization of periodic elastic composites and locally resonant sonic
  materials. {\em Phys. Rev. B\/} {\bf 83}, 104103.

\bibitem[Nevard \& Keller(1997)]{nevard97a}
{\sc Nevard, J. \& Keller, J.~B.} 1997 Homogenization of rough boundaries and
  interfaces. {\em SIAM J. Appl. Math.\/} {\bf 57}, 1660--1686.

\bibitem[Nolde {\em et~al.\/}(2011)Nolde, Craster \& Kaplunov]{nolde11a}
{\sc Nolde, E., Craster, R.~V. \& Kaplunov, J.} 2011 High frequency
  homogenization for structural mechanics. {\em J. Mech. Phys. Solids\/} {\bf
  59}, 651--671.

\bibitem[Panasenko(2005)]{panasenko05a}
{\sc Panasenko, G.} 2005 {\em Multi-scale modelling for structures and
  composites\/}. Dordrecht: Springer.

\bibitem[Pendry {\em et~al.\/}(1999)Pendry, Holden, Stewart \&
  Youngs]{pendry99a}
{\sc Pendry, J.~B., Holden, A.~J., Stewart, W.~J. \& Youngs, I.} 1999 Magnetism
  from conductors and enhanced nonlinear phenomena. {\em IEEE Trans. Micr.
  Theo. Tech.\/} {\bf 47}, 2075.

\bibitem[Pendry {\em et~al.\/}(2004)Pendry, Martin-Moreno \&
  Garcia-Vidal]{pendry04a}
{\sc Pendry, J.~B., Martin-Moreno, L. \& Garcia-Vidal, F.~J.} 2004 Mimicking
  surface plasmons with structured surfaces. {\em Science\/} {\bf 305},
  847--848.

\bibitem[Petit(1980)]{petit80a}
{\sc Petit, R.} 1980 {\em Electromagnetic theory of gratings, Topics in current
  physics\/}. Springer-Verlag, Berlin.

\bibitem[Popov(2012)]{popov12a}
{\sc Popov, E.} 2012 {\em Gratings: Theory and Numerical Applications\/}.
  Aix-Marseille University.

\bibitem[Porter \& Evans(1999)]{porter99a}
{\sc Porter, R. \& Evans, D.~V.} 1999 Rayleigh-{B}loch surface waves along
  periodic gratings and their connection with trapped modes in waveguides. {\em
  J. Fluid Mech.\/} {\bf 386}, 233--258.

\bibitem[Ramakrishna(2005)]{ramakrishna05a}
{\sc Ramakrishna, S.~A.} 2005 Physics of negative refractive index materials.
  {\em Rep. Prog. Phys.\/} {\bf 68}, 449--521.

\bibitem[Sanchez-{P}alencia(1980)]{sanchez80a}
{\sc Sanchez-{P}alencia, E.} 1980 {\em Non-homogeneous media and vibration
  theory\/}. Berlin: Springer-Verlag.

\bibitem[Sengupta(1959)]{sengupta59a}
{\sc Sengupta, D.} 1959 On the phase velocity of wave propagation along an
  infinite {Y}agi structure. {\em IRE Trans. Antennas Propagat\/} {\bf 7},
  234--239.

\bibitem[Wilcox(1984)]{wilcox84a}
{\sc Wilcox, C.~H.} 1984 {\em Scattering theory for diffraction gratings\/}.
  Springer-Verlag.

\end{thebibliography}
\end{document}